\documentclass[twocolumn,aps,prc,showpacs,superscriptaddress,floatfix,nofootinbib]{revtex4-2}

\usepackage{color}
\usepackage{url}
\usepackage{amsmath}

\usepackage{enumitem}

\usepackage{hyperref}

\makeatletter
\renewcommand{\p@subsection}{}
\renewcommand{\p@subsubsection}{}
\makeatother

\usepackage{lineno}

\usepackage{multirow}

\newcommand*{\myvcell}[1]{%
\begingroup
   \renewcommand*{\arraystretch}{0.0}%
   \begin{tabular}[c]{@{}>{\raggedright\arraybackslash}p{\linewidth}@{}}#1\end{tabular}%
\endgroup
}

\usepackage{array}
\newcolumntype{L}[1]{>{\raggedright\let\newline\\\arraybackslash\hspace{0pt}}m{#1}}
\newcolumntype{C}[1]{>{\centering\let\newline\\\arraybackslash\hspace{0pt}}m{#1}}
\newcolumntype{R}[1]{>{\raggedleft\let\newline\\\arraybackslash\hspace{0pt}}m{#1}}

\usepackage{rotating}
\usepackage{epsfig} 
\usepackage{graphicx}
\usepackage[caption=false]{subfig}

\graphicspath{ {./} {./plot/}, {./image/} }
\DeclareGraphicsExtensions{ .pdf, .png}

\usepackage{xcolor}
\usepackage{xparse,xcoffins}

\ExplSyntaxOn
\NewCoffin\imagecoffin
\NewCoffin\labelcoffin

\keys_define:nn { fqwang/label }
 {
  label   .tl_set:N = \l_fqwang_label_tl,
  labelbox .bool_set:N = \l_fqwang_label_box_bool,
  labelbox .default:n = true,
  fontsize .tl_set:N = \l_fqwang_label_size_tl,
  fontsize .initial:n = \normalsize,
  pos .choice:,
  pos/nw .code:n = \tl_set:Nn \l_fqwang_label_pos_tl { left,up },
  pos/ne .code:n = \tl_set:Nn \l_fqwang_label_pos_tl { right,up },
  pos/sw .code:n = \tl_set:Nn \l_fqwang_label_pos_tl { left,down },
  pos/se .code:n = \tl_set:Nn \l_fqwang_label_pos_tl { right,down },
  pos/nwhigh .code:n = \tl_set:Nn \l_fqwang_label_pos_tl { left,uphigh },
  pos/nehigh .code:n = \tl_set:Nn \l_fqwang_label_pos_tl { right,uphigh },
  pos/swhigh .code:n = \tl_set:Nn \l_fqwang_label_pos_tl { left,downhigh },
  pos/sehigh .code:n = \tl_set:Nn \l_fqwang_label_pos_tl { right,downhigh },
  pos/nwlow .code:n = \tl_set:Nn \l_fqwang_label_pos_tl { left,uplow },
  pos/nelow .code:n = \tl_set:Nn \l_fqwang_label_pos_tl { right,uplow },
  pos/swlow .code:n = \tl_set:Nn \l_fqwang_label_pos_tl { left,downlow },
  pos/selow .code:n = \tl_set:Nn \l_fqwang_label_pos_tl { right,downlow },
  pos/n .code:n = \tl_set:Nn \l_fqwang_label_pos_tl { hc,up },
  pos/w .code:n = \tl_set:Nn \l_fqwang_label_pos_tl { left,vc },
  pos/s .code:n = \tl_set:Nn \l_fqwang_label_pos_tl { hc,down },
  pos/e .code:n = \tl_set:Nn \l_fqwang_label_pos_tl { right,vc },
  pos .initial:n = nw,
  unknown .code:n   = \clist_put_right:Nx \l_fqwang_label_clist
                       { \l_keys_key_tl = \exp_not:n { #1 } }
 }

\clist_new:N \l_fqwang_label_clist
\box_new:N \l_fqwang_label_box
\box_new:N \l_fqwang_label_image_box

\NewDocumentCommand{\xincludegraphics}{O{}m}
 {
  \group_begin:
  \tl_clear:N \l_fqwang_label_tl
  \clist_clear:N \l_fqwang_label_clist
  \keys_set:nn { fqwang/label } { #1 }
  \tl_if_empty:NTF \l_fqwang_label_tl
   {
    \fqwang_includegraphics:Vn \l_fqwang_label_clist { #2 }
   }
   {
    \SetHorizontalCoffin\imagecoffin
     {
      \fqwang_includegraphics:Vn \l_fqwang_label_clist { #2 }
     }
    \SetHorizontalCoffin\labelcoffin
     {
      \raisebox{\depth}
       {
        \bool_if:NTF \l_fqwang_label_box_bool
         { \fcolorbox{white}{white}{\l_fqwang_label_size_tl\l_fqwang_label_tl} }
         { \l_fqwang_label_size_tl\l_fqwang_label_tl }
       }
     }
    \SetVerticalPole\imagecoffin{left}{36pt+\CoffinWidth\labelcoffin/2}
    \SetVerticalPole\imagecoffin{right}{\Width-36pt-\CoffinWidth\labelcoffin/2}
    \SetHorizontalPole\imagecoffin{up}{\Height-12pt-\CoffinHeight\labelcoffin/2}
    \SetHorizontalPole\imagecoffin{down}{12pt+\CoffinHeight\labelcoffin/2}
    \SetHorizontalPole\imagecoffin{uphigh}{\Height-6pt-\CoffinHeight\labelcoffin/2}
    \SetHorizontalPole\imagecoffin{downlow}{6pt+\CoffinHeight\labelcoffin/2}
    \SetHorizontalPole\imagecoffin{uplow}{\Height-18pt-\CoffinHeight\labelcoffin/2}
    \SetHorizontalPole\imagecoffin{downhigh}{18pt+\CoffinHeight\labelcoffin/2}
    \use:x{\JoinCoffins\imagecoffin[\l_fqwang_label_pos_tl]\labelcoffin[vc,hc]} 
    \TypesetCoffin\imagecoffin
   }
   \group_end:
 }
\NewDocumentCommand{\setlabel}{m}
 {
  \keys_set:nn { fqwang/label } { #1 }
 }
\cs_new_protected:Nn \fqwang_includegraphics:nn
 {
  \includegraphics[#1]{#2}
 }
\cs_generate_variant:Nn \fqwang_includegraphics:nn { V }

\ExplSyntaxOff

\begin{document}

\newcommand{\snn}{\sqrt{s_{\textsc{nn}}}}
\newcommand{\hijing}{{\textsc{hijing}}}
\newcommand{\zdc}{\{{\textsc{zdc}}\}}
\newcommand{\pp}{$pp$}
\newcommand{\pA}{$p$A}
\newcommand{\dA}{$d$A}
\newcommand{\AuAu}{Au+Au}
\newcommand{\Nch}{N_{\rm ch}}
\newcommand{\pt}{p_T}
\newcommand{\dphi}{\Delta\phi}
\newcommand{\deta}{\Delta\eta}
\newcommand{\pair}{{\rm pair}}
\newcommand{\low}{{\rm low}}
\newcommand{\high}{{\rm high}}
\newcommand{\sub}{{\rm sub}}
\newcommand{\two}{\{2\}}
\newcommand{\four}{\{4\}}

\newcommand {\mean}[1]   {\langle{#1}\rangle}

\newcommand {\red}[1]   {\textcolor{red}{#1}}
\newcommand {\blue}[1]  {\textcolor{blue}{#1}}
\newcommand {\green}[1] {\textcolor{green}{#1}}
\newcommand {\pink}[1] {\textcolor{magenta}{#1}}

\title{Review of nonflow estimation methods and uncertainties in relativistic heavy-ion collisions}

\author{Yicheng Feng}
\email{feng216@purdue.edu}
\address{Department of Physics and Astronomy, Purdue University, West Lafayette, IN 47907}
\author{Fuqiang Wang}
\email{fqwang@purdue.edu}
\address{Department of Physics and Astronomy, Purdue University, West Lafayette, IN 47907}


\begin{abstract}
    Collective anisotropic flow, where particles are correlated over the entire event, is a prominent phenomenon in relativistic heavy-ion collisions and is sensitive to the properties of the matter created in those collisions. It is often measured by two- and multi-particle correlations and is therefore  contaminated by nonflow, those genuine few-body correlations unrelated to the global event-wise correlations. Many methods have been devised to estimate nonflow contamination with various degrees of successes and difficulties. Here, we review those methods pedagogically, discussing the pros and cons of each method, and give examples of ballpark estimate of nonflow contamination and associated uncertainties in relativistic heavy-ion collisions. We hope such a review of the various nonflow estimation methods in a single place would prove helpful to future researches.
\end{abstract}


\maketitle

\section{Introduction} \label{sec:introduction}
Collective anisotropic flow is a hallmark of heavy-ion (nucleus-nucleus) collisions~\cite{Reisdorf:1997fx}. It refers to azimuthal angular correlations among particles over the entire event -- every particle is correlated with every other particle. 
One specific mechanism that gives rise to such global correlations (collective effects) is 
hydrodynamic flow, where the initial geometry anisotropy in finite impact parameter collisions is converted by interactions into momentum space anisotropy of particles~\cite{Ollitrault:1992bk,Teaney:2000cw,Kolb:2003dz}. 
All particles are thus correlated in azimuth to the ellipse-shaped collision geometry, e.g.~with respect to the impact parameter direction. The anisotropy is nonzero even in head-on (zero impact parameter) collisions because of position fluctuations of nucleons inside the colliding nuclei giving rise to finite eccentricities~\cite{Alver:2008zza,Alver:2010gr}, which defines the so-called participant plane azimuth. Because of the same reason, nonzero anisotropy can also emerge in small-system collisions, such as proton-proton (\pp), proton-nucleus (pA), deuteron-nucleus (dA), and helium-nucleus collisions~\cite{Li:2012hc,Nagle:2018nvi}. In fact, for any impact parameter, it is still the participant plane direction that is most relevant for azimuthal anisotropic flow, whose departure from that with respect to the impact parameter direction constitutes flow fluctuations, stemming out of fluctuations of the participant plane direction about the impact parameter direction~\cite{Ollitrault:2009ie,Qiu:2011iv}.

The interactions in the system created in relativistic heavy-ion collisions at RHIC (Relativistic Heavy-Ion Collider) and the LHC (Large Hadron Collider), presumably the quark-gluon plasma (QGP), are governed by quantum chromodynamics (QCD)~\cite{Adcox:2004mh,Adams:2005dq,Arsene:2004fa,Back:2004je,Muller:2012zq,Roland:2014jsa}. 
It is generally believed that ultra-strong interactions are required to produce the observed large anisotropy (or flow) in heavy-ion collisions, and the QGP created in those collisions is a nearly perfect fluid~\cite{Gyulassy:2004zy} and can be well described by hydrodynamics with little viscosity~\cite{Romatschke:2007mq,Song:2010mg,Heinz:2013th}. In peripheral heavy-ion collisions and small-system collisions, the interactions may not be intense enough where hydrodynamics could be applicable and the escape mechanism may be at work~\cite{He:2015hfa,Romatschke:2015dha,Kurkela:2018qeb,Kurkela:2021ctp}. 
Recently, it has been shown that collective effects can be generated with few interactions~\cite{Kurkela:2018qeb} and hydrodynamics is an attractor for the behaviour of many-body systems~\cite{Brewer:2022ifw,Jankowski:2023fdz,Boguslavski:2023jvg,Rajagopal:2024lou}.
It is important to keep in mind that the word ``flow'', while suggestive, does not necessarily mean hydrodynamic flow.

The collision geometry--the impact parameter vector of a heavy-ion collision or generally the geometric shape of the interaction zone--is experimentally unknown. It is often reconstructed as a proxy from the final-state particle azimuthal distribution as the symmetry harmonic plane~\cite{Poskanzer:1998yz}. The particle azimuthal distribution can be written in the Fourier series~\cite{Voloshin:1994mz}:
\begin{equation}
    \frac{dN}{d\phi} = \frac{N}{2\pi}\left(1+2\sum_{n=1}^\infty v_n\cos n(\phi-\psi_n)\right)\,,
    \label{eq:fourier}
\end{equation}
where $\phi$ is the particle azimuthal angle and $\psi_n$ is that of the $n^{\rm th}$ order harmonic plane.
The anisotropic harmonic flow is then given by
\begin{equation}
    v_n=\mean{\cos n(\phi-\psi_n)}\,.
    \label{eq:vn}
\end{equation}
Here $n=1,2,3,...$; $v_1$ is called directed flow, $v_2$ elliptic flow, $v_3$ triangular flow, and so on.
Because $\psi_n$ is reconstructed from particles, the $v_n$ values in Eq.~(\ref{eq:vn}) effectively measure two-particle correlation strengths. Namely, it can be obtained  from two-particle cumulant, or the Fourier coefficient $V_n$ of the two-particle correlation function,
\begin{equation}
    \frac{dN_\pair}{d\dphi} = \frac{N_\pair}{2\pi}\left(1+2\sum_{n=1}^\infty V_n\cos n\dphi\right)\,,
    \label{eq:pair}
\end{equation}
where $\dphi=\phi_1-\phi_2$ is the azimuthal angle difference of a particle pair.
The Fourier coefficient is simply
\begin{equation}
    V_n = \mean{\cos n\dphi} \equiv c_n\,,
    \label{eq:Vn}
\end{equation}
where $c_n$ is also used in literature~\cite{STAR:2022pfn,STAR:2023wmd}.
Under the presence of only collective flow, it follows straightforwardly that
\begin{equation}
    V_n = v_n^2\two\,,
    \label{eq:factorization}
\end{equation}
where $v_2\two$ denotes two-particle cumulant flow~\cite{Borghini:2000sa,Borghini:2001vi,Borghini:2001zr,Bilandzic:2010jr}.

If measured by the two-particle cumulant method of Eqs.~(\ref{eq:pair},\ref{eq:Vn}), the $V_n$ is not necessarily related to the collision geometry. For example, the gluon field from each incoming nucleus can be correlated and this correlation may result in a final-state anisotropy $V_n$~\cite{Dusling:2012iga,Dusling:2015gta}. Such an initial-state correlation permeates over an entire collision event, so it is flow, but not part of hydrodynamic flow.

\begin{figure}[hbt]
    \includegraphics[width=1.0\linewidth]{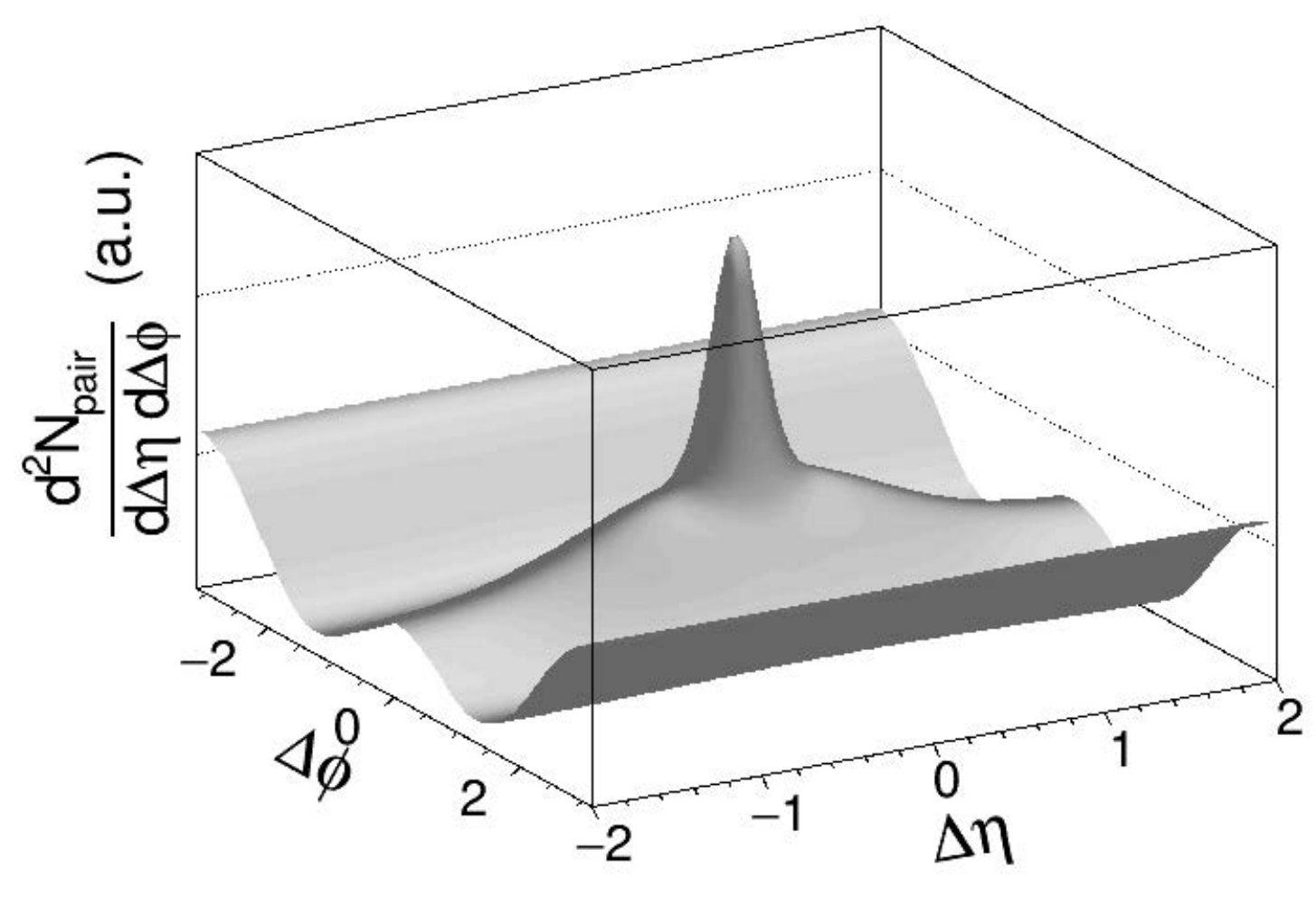}
    \caption{Illustration of nonflow correlations in two-particle $(\deta,\dphi)$ angular distribution. The near-side peak at $\dphi\sim0$ is mainly short ranged ($\deta\sim0$). There could be various near-side short-range correlations: HBT is very short ranged (sharp peak), resonance decay and intra-jet correlations are short ranged with a $\deta$ width on the order of one unit. The away-side ridge at $\dphi\sim\pi$ is long ranged and comes mainly from dijet correlations and global momentum conservation. Underlying these nonflow peaks are the majority pair distribution (zero-suppressed), modulated by anisotropic flows of various harmonic orders in $\dphi$, which could have dependencies on $\eta$ (on single particle level) and $\deta$ (on two-particle level, referred to as flow decorrelation).}
    \label{fig:2D}
\end{figure}

There is, however, an important contribution to $V_n$ (or more generally, any variables) measured via correlations, and that is nonflow: two- and few-particle genuine correlations that have nothing to do with the collision geometry or event-wise global correlations~\cite{Borghini:2000cm,Borghini:2002mv,Wang:2008gp,Ollitrault:2009ie,Alver:2010rt}. Nonflow includes all other contributions to the measured $V_n$ except the collective flow. 
Examples of nonflow correlations are those between daughter particles from a resonance decay, particles from a jet shower originated from an energetic parton (quark or gluon)~\cite{Jacobs:2004qv,Wang:2013qca}, hadrons from string fragmentation~\cite{Andersson:1983jt}, Hanbury-Brown Twiss (HBT) interferometry~\cite{Lisa:2005dd}, and global momentum conservation~\cite{Borghini:2006yk}. 
Figure~\ref{fig:2D} illustrates those correlations in a two-particle angular correlation plot in $(\deta,\dphi)$ where $\deta=\eta_1-\eta_2$ is the two-particle pseudorapidity difference.
Most of the nonflow correlations (HBT, resonance decays, intra-jet correlations, string fragmentation) are short-ranged (small-angle) correlations contributing to the near-side ($\dphi\sim 0$) peak at $\deta\sim 0$. HBT is relevant at very small angle differences and its effect is generally  minor because of the limited phase space at small angles. There is a long-range contribution from dijet correlations contributing to the away-side ridge at $\dphi\sim\pi$ that has little $\deta$ dependence (because of the stochastic sampling of kinematics of the underlying parton-parton scattering producing the dijet)~\cite{Baier:2000mf,Jacobs:2004qv}.
It is noteworthy that the nonflow contributions from jets are those intra- and inter-jet hadron correlations, whereas the azimuthal anisotropy of jet-axis orientations, a result of path-length dependent partonic energy loss or jet quenching~\cite{Wang:1991xy,Wang:1998bha}, is related to the collision geometry and is thus part of flow. 
Global momentum conservation contributes to the away-side ridge, but only to the first harmonic~\cite{Borghini:2006yk}.

All those nonflow correlations contribute to the $V_n$ of Eq.~(\ref{eq:Vn}), so the factorization in Eq.~(\ref{eq:factorization}) is no longer valid~\cite{Ollitrault:2009ie,Zhu:2005qa,Kikola:2011tu,ALICE:2011svq,ALICE:2017xzf}.
A keen interest in relativistic heavy-ion collisions is to measure collective anisotropic flows arising from final-state interactions to probe the properties of the QGP by, for example, comparing to hydrodynamic calculations~\cite{Huovinen:2006jp,Shuryak:2008eq,Heinz:2013th,Gale:2013da}.
Thus, the goal is to measure the collective anisotropic flows from global event-wise correlations related to the collision geometry. To this end, nonflow contamination in $V_n$ must first be subtracted with faithful systematic uncertainties~\cite{Wang:2024zml}.
Recently, azimuthal anisotropies similar to those in heavy-ion collisions have been observed in small-system p/d/$^3$He+A and pp collisions~\cite{Khachatryan:2010gv,CMS:2012qk,CMS:2013jlh,Abelev:2012ola,ABELEV:2013wsa,Aad:2012gla,ATLAS:2016yzd,PHENIX:2018lia,PHENIX:2021ubk,Adamczyk:2015xjc,STAR:2022pfn,STAR:2023wmd}, prompting intense interest in possible collective phenomenon in small systems. 
There, nonflow correlations are dominating, if not the entirety of, the observed anisotropies. The challenge to remove nonflow correlations is particularly serious in order to identify any collective flow signals in small systems~\cite{Lim:2019cys,Nagle:2021rep,Abdulhamid:2024ifx}.

Many experimental methods have been devised to estimate and/or subtract nonflow contributions from azimuthal anisotropy measurements. This note aims to give a pedagogical review of those methods in a single place, hopefully useful for future researches.
The possible contributions from geometry-unrelated initial-state gluon correlations~\cite{Dusling:2012iga,Dusling:2015gta} need also to be considered. This is however outside the scope of this note.

It should be noted that nonflow is an issue mostly relevant to two-particle cumulant measurements as in Eqs~(\ref{eq:pair},\ref{eq:Vn},\ref{eq:factorization}). Nonflow is significantly reduced in multiparticle cumulants as clusters composed of several particles are sparse~\cite{STAR:2002hbo,STAR:2004jwm,ALICE:2014dwt,CMS:2015yux,ALICE:2019zfl,ATLAS:2017rtr}. Moreover, their contributions are diluted by multiplicity to higher powers~\cite{Borghini:2000sa,Borghini:2001vi}. 
The flow fluctuation effects are, however, different in $v_2\four$ and $v_2\two$, negative in the former and positive in the latter~\cite{Ollitrault:2009ie,Voloshin:2007pc}. For physics pertinent to two-particle correlations, such as the chiral magnetic effect~\cite{Kharzeev:2015znc,Zhao:2019hta}, measurement of $v_2\two$ is crucial and that of $v_2\four$ is largely irrelevant. Furthermore, for rare probes, low-multiplicity events, or higher-order harmonics, $v_n\four$ measurements are insurmountable statistically, so $v_n\two$ measurements are essential. Because of all these reasons, nonflow studies in two-particle cumulant measurements are indispensable. 

\section{Nonflow estimation methods}
There are three main categories of nonflow estimation methods: 
(1) applying pseudorapidity $\deta$ gaps between particle pair, 
(2) subtracting correlations from low-multiplicity events, and 
(3) performing data-driven fits to two-particle correlation functions.

\subsection{$\deta$-gap Methods}\label{sec:deta}
Nonflow correlations are primarily short ranged in $\deta$. 
Those short-range nonflow correlations can be suppressed by applying a $\deta$ gap between particle pairs used in the correlation analysis.
The $\deta$ gap method cannot suppress long-range nonflow correlations.

\subsubsection{Simple $\deta$-gap method}\label{sec:simple}
The simplest way to suppress nonflow is to impose a $\deta$ gap between the two hadrons in two-particle cumulant measurement of $V_n$~\cite{ALICE:2011ab,ATLAS:2011ah,ALICE:2016kpq,STAR:2008ftz,CMS:2012xss,CMS:2017xgk,PHENIX:2018lia}. This method is simple and straightforward to implement in data analysis. The shortcomings are obvious: 
\begin{enumerate}
    \item[i)] the method is not clean--how much nonflow is eliminated depends on the $\deta$-gap size relative to the width of the near-side short-range correlations;
    \item[ii)] away-side jet correlations cannot be eliminated because dijets can be widely separated in $\eta$, and how much they contribute to $V_n$ is unknown a priori;
    \item[iii)] the $\eta$-dependencies~\cite{Back:2004mh,STAR:2004jwm,ALICE:2016tlx,CMS:2016est} and longitudinal decorrelations~\cite{Bozek:2010vz,Xiao:2012uw,CMS:2015xmx,ATLAS:2017rij} of flow and flow fluctuations would yield different measurements of $V_n$ with different $\deta$-gap sizes. 
\end{enumerate}
The last is not related to nonflow, but flow and flow fluctuations. However, since they cannot be distinguished by a $\deta$-gap analysis, the effects of flow and nonflow are mixed. These shortcomings make the uncertainty estimation of a particular $\deta$-gap result difficult.

\subsubsection{Two-subevent method}
Instead of applying $\deta$-gap between the two particles, one may use the two-subevent method where one particle is taken from one subevent in a given $\eta$ region and the other from another subevent in a different $\eta$ region, and the two subevents are separated in $\eta$ with a certain $\deta$ gap~\cite{STAR:2000ekf,STAR:2008ftz,STAR:2011gzz,ATLAS:2014txd,Jia:2017hbm,CMS:2019lin}. This has the advantage that the cumulant calculation can be applied which involves only single-particle loops~\cite{Bilandzic:2010jr,Bilandzic:2013kga}, thus saving tremendous computing time compared to double loops of particle pairs. Obviously, all the shortcomings of the simple $\deta$-gap method described in Sect.~\ref{sec:simple} are present in the two-subevent method.

\subsubsection{Three- and four-subevent methods}

\begin{figure}[hbt]
    \includegraphics[width=\linewidth]{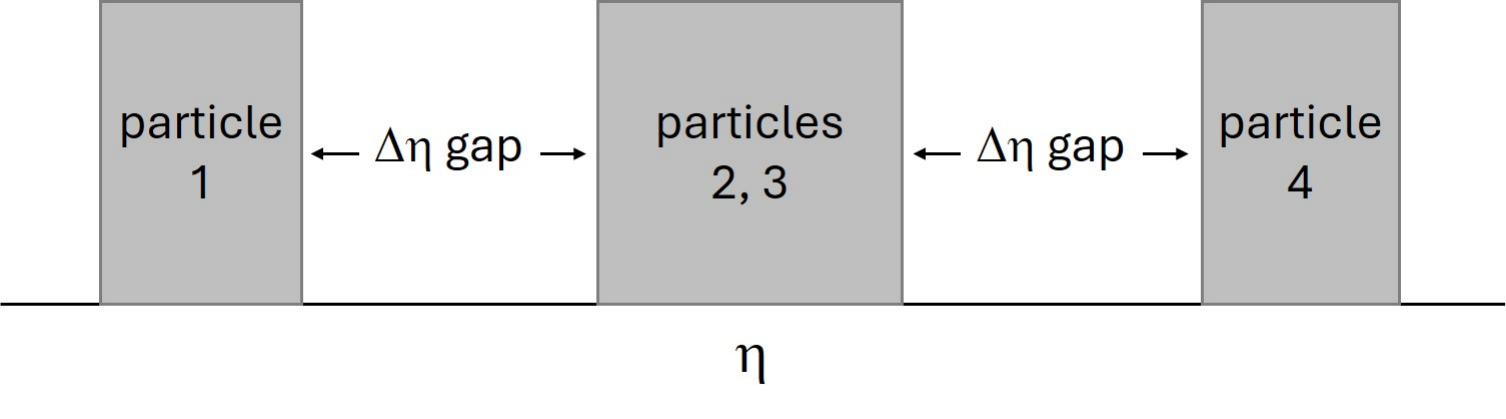}
    \caption{Sketch of the three-subevent method. Four-particle cumulant is formed by taking two particles from one subevent and one particle from each of the other two subevents. All nonflow correlations are suppressed including back-to-back dijet correlations provided sufficient $\deta$ gaps, except global momentum conservation.}
    \label{fig:sub}
\end{figure}

The away-side jet correlations are long ranged in $\deta$, and thus the simple $\deta$-gap method and the two-subevent method cannot suppress the away-side nonflow correlations.
One can use three- and four-subevent method to calculate multi-particle cumulants to suppress the away-side nonflow correlations from dijets by separating those subevents well in $\eta$~\cite{DiFrancesco:2016srj,Jia:2017hbm,ATLAS:2017rtr}. 
Figure~\ref{fig:sub} illustrates the idea behind the three-subevent method of four-particle cumulant, where dijets appear in at most two subevents and the third subevent is free of nonflow dijet correlations once the $\deta$-gaps are made wide enough. As a result, the correlations among the particles--two from one subevent and one from each of the other two subevents--are due only to collective flow. 
The four-subevent method for four-particle cumulant is similar, except that one particle is taken from each of the four subevents.

The multi-subevent method, with sufficient $\deta$ gaps, should suppress essentially  all nonflow correlations (except those in $V_1$ from global momentum conservation~\cite{Borghini:2006yk}). This method also takes advantage of cumulant calculations~\cite{Bilandzic:2010jr,Bilandzic:2013kga}, greatly reducing the computing demand. 
The shortcomings are as same as those from the simple $\deta$-gap method described in Sect.~\ref{sec:simple}, except (ii) provided sufficiently large $\deta$ gaps.

\subsection{Low-multiplicity subtraction methods}\label{sec:low}

The general idea behind this category of methods is that particle correlations in peripheral and low-multiplicity collisions are primarily due to nonflow, and
those in central and high-multiplicity collisions come from collective flow with some contamination from nonflow correlations.
Nonflow in $V_n$, being due to few-body correlations, is diluted by combinatorial particle pairs, and is therefore significantly reduced by high multiplicities.  
As a result, nonflow dominates in peripheral/low-multiplicity collisions and any flow may be neglected, and in central/high-multiplicity collisions flow may prevail. 
The degree of nonflow contamination at high multiplicities may then be assessed from particle correlation measurements in low-multiplicity events. 
The key questions are:
\begin{itemize}
    \item[i)] at what low multiplicity the correlations can be considered all as nonflow, and
    \item[ii)] how this low-multiplicity nonflow can be modeled or utilized to gauge nonflow in high-multiplicity/central events.
\end{itemize}

\subsubsection{Inverse multiplicity scaling method}\label{sec:scaling}
Resonance abundances are expected to scale approximately with the final-state multiplicity~\cite{Braun-Munzinger:1994ewq,Andronic:2017pug}. If the scaling is exact and if resonance kinematic distributions, and thus the average correlations between decay daughters, do not change with event multiplicity, then nonflow contribution from resonance decays to $V_n$ by Eq.~(\ref{eq:Vn}) is inversely proportional to  multiplicity ($N$). 
Take the number of particles to be $N_\low$ in low-multiplicity events and $N_\high$ in high-multiplicity events, and take the number of pairs to be $N^2$ (valid under Poisson statistics), then
\begin{eqnarray}
    N^2_\high V_n ^\high &=& N^2_\high V_n^\sub + N^2_\low V_n ^\low\cdot\frac{N_\high}{N_\low}\,, \label{eq:scaling1} \\
    V_n ^\sub &=& V_n ^\high - \frac{N_\low}{N_\high}V_n ^\low\,. \label{eq:scaling}
\end{eqnarray}
Here, it is assumed that the number of nonflow ``sources'' scales with multiplicity, leading to the factor $N_\high/N_\low$ in Eq.~(\ref{eq:scaling1}). 
The notation $V_n ^\sub$ denotes low-multiplicity subtracted $V_n $ with the ultimate goal to be the nonflow-subtracted anisotropic flow.
This method is sometimes called the $c_0$ method where $c_0\equiv N$~\cite{STAR:2022pfn,STAR:2023wmd}. 

The $1/N$ scaling implies not only a scaling in the abundance of nonflow sources, but also that the physics of nonflow correlations does not change from low- to high-multiplicity collisions. Neither is good assumption for nonflow contribution from jet correlations:
\begin{itemize}
    \item Jets are produced by hard processes whose abundance increases with $N$ more strongly than linearly in heavy-ion collisions~\cite{Wang:1991xy,Wang:1998bha}. The $1/N$ scaling would be an underestimate of this part of nonflow. However, the majority nonflow contribution comes from relatively low transverse momentum ($\pt$) jets or minijets, whose production may be more closely proportional to $N$.
    \item Jets are modified by the nuclear medium created in relativistic heavy-ion collisions. Such modifications broaden/suppress the angular correlations between jet fragments~\cite{Adams:2005ph,Vitev:2005yg,STAR:2011ryj,Agakishiev:2014ada,STAR:2010uoq,Wang:2013qca,Mueller:2016xoc}, making nonflow weaker.
    \item On the other hand, jets lose energy via collisional and radiative partonic energy loss~\cite{Baier:2000mf,Gyulassy:2003mc,Qin:2007rn}, resulting in more lower $\pt$ particles and particle pairs, and thus stronger nonflow correlations. 
\end{itemize}
Quantitatively, these effects are $\pt$ dependent, and the interplay among them determines the final dependency of nonflow on collision centrality/multiplicity. 
Simulations by the \hijing\ model--a jet production model without hydrodynamic flow so its correlations are entirely of nonflow nature\footnote{The pathlength dependent energy loss results in an anisotropy of jet fragments relative to the overall collision geometry, thus flow. This flow is encoded in the correlations between jet fragments and underlying particles in \hijing\ but is negligible compared to the nonflow caused by intra- and inter-jet correlations~\cite{Zhao:2019kyk}.}--indicate only a modest excess (10--20\%) of correlation strengths in central \AuAu\ collisions  at $\snn=200$~GeV than the $1/N$ scaling from peripheral collisions would entail~\cite{Wang:2024zml}. Nevertheless, this is one inferior part of the method.

Besides the strong assumptions, 
the $1/N$ scaling method also explicitly assumes 
that the correlations in low-multiplicity collisions are all nonflow. The flow result in high-multiplicity events extracted from this method will thus inevitably depend on what low-multiplicity events are considered as the pure-nonflow baseline. 
The selection of very low-multiplicity or very peripheral events seems to be the natural choice, but they could be biased towards too soft underlying nucleon-nucleon scatterings or diffractive interactions. This is a selection bias 
and has indeed been observed in small systems at RHIC~\cite{PHENIX:2013jxf,Adamczyk:2014fcx} and in peripheral Pb+Pb collisions at the LHC~\cite{ALICE:2018ekf}. Such effects at the LHC start to become important at 80\% centrality mark towards more peripheral collisions~\cite{ALICE:2018ekf}. The nonflow in those events may therefore not be a good reflection of nonflow in central collisions; in other words, the lowest multiplicity collisions are unnecessarily the best baseline for nonflow subtraction.

This issue is more severe in analyzing anisotropies in small-system collisions, such as \pp, \pA, and \dA\ collisions. Because of multiplicity selection biases~\cite{PHENIX:2013jxf,Adamczyk:2014fcx}, the low- and high-multiplicity collisions of those small systems can be vastly different in physics, including nonflow correlations. 
High-multiplicity \pp\ events are likely  biased towards jet production, whereas low-multiplicity events are likely biased towards softer-than-average interactions. As a result, the $1/N$ scaling may completely fail in gauging nonflow contamination in small systems. %


Multiplicity selection bias is likely insignificant in central heavy-ion collisions--there are many underlying nucleon-nucleon interactions. For example, it is highly unlikely to have all those interactions to produce jets.\footnote{This is of course valid only up to a certain high-multiplicity limit. When one demands high multiplicity with extremely small cross section,  multiplicity selection biases would also manifest in central heavy-ion collisions.} Therefore, minimum-bias \pp\ collisions would be the best baseline at our disposal for nonflow subtraction for central heavy-ion collisions within the scope of the $1/N$ scaling method.

In Eq.~(\ref{eq:scaling1}), it is implicitly assumed, by the $N_\high^2$ in front of $V_n^\sub$, that those nonflow particles have attained the global collective flow with all other particles in central events, while still retaining their genuine few-body correlations. This can happen in two scenarios: the particles have attained collective flow from system expansion but still retaining their correlations from early times, or the correlations are generated late in time after the collective flow has built up. 
These are reasonable for heavy-ion collisions but may not be justified in small-system collisions. It is possible that those nonflow particles, such as jets produced early, have exited the collision zone without participating in the final-state interactions that are responsible for the generation of flow. In such a case, the $V_n$ estimated by Eq.~(\ref{eq:scaling}) would be an underestimate of flow of the underlying event in those high-multiplicity small-system collisions, and that from the template fit method (to be discussed in Sect.~\ref{sec:template}) may be a better reflection of the underlying flow. 
Another good example is resonance decays, where any flow attained by the parent particle is inherited in the decay daughters while the nonflow correlations between the daughter particles (dependent of the decay kinematics) still contribute to the final-state two-particle anisotropy. This applies to both heavy-ion and small-system collisions, and therefore it makes more sense to count all particles, including those decay products, in the baseline to evaluate azimuthal anisotropies. How accurate  this part of nonflow is estimated will of course depend on how the abundances and kinematic distributions of resonances vary with the event multiplicity.

To recap, the advantage of the $1/N$ scaling method is that it is simple to implement. The shortcomings are the inherent  assumptions, namely, 
\begin{itemize}
    \item[i)] it assumes that all correlations are nonflow in low-multiplicity events, which begs the question how low in multiplicity is a good baseline;
    \item[ii)] it assumes the $1/N$ scaling of nonflow correlations, implying no change in the physics of nonflow correlations from low- to high-multiplicity collisions, a strict proportionality to $N$ of the abundance of nonflow sources, and no selection biases for those low- (and high-)multiplicity events;
    \item[iii)] it assumes that those nonflow particles, while still genuinely correlated among themselves, have also attained individually the same collective flow as the rest of the collision event.
\end{itemize}
The strong assumption of ii) and the fact that it is a priori unknown how nonflow correlations vary with the collision centrality/multiplicity make it difficult to assess the robustness and the associated uncertainties of the estimated nonflow by the simple $1/N$ scaling method.

\subsubsection{Near-side jetlike yield-scaled subtraction}\label{sec:near}
As noted in the Sect.~\ref{sec:simple}, nuclear effects, like jet quenching, result in modifications to jet correlations~\cite{Jacobs:2004qv,Wang:2013qca}, thus nonflow effects change from peripheral to central collisions. 
Likewise, particle production mechanisms and distributions change with centrality, yielding, for example, the baryon-over-meson enhancement from peripheral to central collisions~\cite{PHENIX:2003tvk,STAR:2006uve,ALICE:2013cdo}, which would modify the nonflow effects from resonance decays. 
Any of those changes will cause nonflow to deviate from the simple $1/N$ scaling.

Both jetlike correlations and resonance decays produce a near-side correlation peak at $(\deta,\dphi)\sim(0,0)$. Dijets contribute to an away-side correlation at $\dphi\sim\pi$ but more or less uniform in $\deta$. One may take the difference between small-$\deta$ (short-range) and large-$\deta$ (long-range) correlations, properly normalized, to arrive at a near-side correlated yield, which is primarily composed of contributions from resonance decays and intra-jet fragments. One can then compare this near-side yield in high-multiplicity events to that in low-multiplicity events. Any difference would indicate modification of nonflow correlations from low- to high-multiplicity collisions and/or multiplicity selection biases~\cite{Abelev:2012ola,Aad:2012gla,PHENIX:2013jxf,Adamczyk:2014fcx,ALICE:2018ekf}, or likely both. One may take the ratio of the correlated near-side yields ($Y_\high/Y_\low$) as a scaling factor to apply onto the low-multiplicity nonflow $V_n^\low$ in Eq.~(\ref{eq:scaling}) to possibly take into account  modifications to nonflow correlations~\cite{CMS:2013jlh,Adamczyk:2014fcx,Adamczyk:2015xjc,CMS:2016fnw}. The estimated flow in high-multiplicity events would then be
\begin{equation}
    V_n^\sub = V_n^\high - \frac{Y_\high}{Y_\low}\frac{N_\low}{N_\high}V_n^\low\,.
    \label{eq:near}
\end{equation}
Such a scaling, while an improvement over the simple $1/N$ scaling, comes with its own issues:
\begin{itemize}
    \item[i)] the near-side nonflow correlation shape may change/broaden, the effect of which is not included by the simple scaling of the correlated yield; 
    \item[ii)] the away-side jet correlated yield unlikely scales with the near-side one because of ``trigger'' biases (for example, surface bias at high $\pt$), and even if it scales, the scaling factor $Y_\high/Y_\low$ includes resonance decays besides near-side jet contribution and is therefore already an incorrect scaling factor, needless to say that the away-side correlation shape can be significantly modified because of large jet-quenching effects;
    \item[iii)] the correlated yield analysis may be difficult at low $\pt$ because of the subtraction of large combinatorial background.
\end{itemize}

The near-side yield-scaled subtraction is usually done at the Fourier coefficient level by Eq.~(\ref{eq:near}). Effectively, it is a subtraction of the scaled correlation function in low-multiplicity events from the correlation function in high-multiplicity events, ignoring any changes in the  correlation shape.
Shape changes in nonflow correlations noted in point i) above can, to some extent, be inspected by the small- and large-$\deta$ correlation difference and can in principle be considered in nonflow subtraction. 
Of course, this difference would be purely nonflow correlations only under the assumption that flow is $\deta$ independent. If flow is $\deta$ dependent,  then it becomes difficult to disentangle the two effects. Moreover, the approximate $\deta$-independent away-side correlations are also subtracted in such a difference, so no information on any  shape change in $\dphi$ correlations is retained. The ultimate approach is to use data-driven fitting methods, which will be discussed in Sect.~\ref{sec:fit}.

Two-particle cumulant flow involves particle pairs.
The nonflow correlation shapes are often studied in terms of per-trigger normalized correlated yield. 
Besides the nonflow correlation shape, the number of trigger particles also matters for flow.
If the number of trigger particles does not scale with event multiplicity, then the $1/N$ scaling will be invalid and this is certainly the case for jet correlations. For most low-$\pt$ flow analysis in heavy-ion collisions, this is not an issue because the trigger particle is the charged hadron itself. 
However, this is an issue for small system collisions like \pp\ because of multiplicity selection biases. The number of minijets is biased and will not be proportional to the event multiplicity. 
This imposes a large uncertainty on small-system flow analysis.

\subsubsection{Template fit}\label{sec:template}
\begin{figure}[hbt]
    \includegraphics[width=\linewidth]{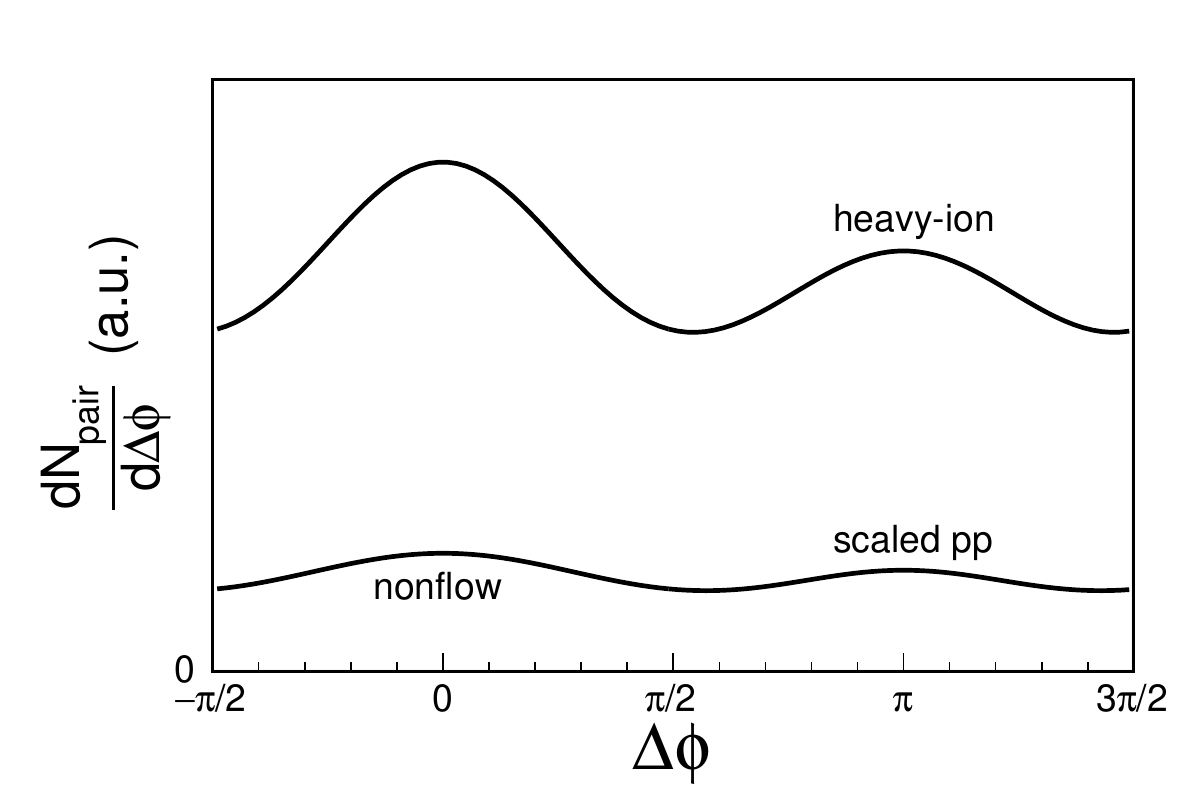}
    \caption{Sketch of the template fit method and the dipole scaling method. The heavy-ion pair correlations are considered to be composed of nonflow correlations scaled from \pp\ and flow correlations. In the template fit, the extracted flow magnitudes are with respect to the pair multiplicity between the two curves, i.e., the nonflow pairs scaled from \pp\ is first subtracted. In the dipole scaling method, the flow magnitudes are with respect to the total pair multiplicity underneath the upper curve. The methods also apply to small system analysis, where the label ``heavy-ion'' is to be replaced by ``high-multiplicity events'' and ``scaled \pp'' by ``low-multiplicity events.''}
    \label{fig:template}
\end{figure}

In the template fit method~\cite{ATLAS:2015hzw,ATLAS:2016yzd,STAR:2022pfn,STAR:2023wmd}, the two-particle correlation is assumed to be composed of a scaled correlation from low-multiplicity events (considered all as nonflow) and a series of Fourier harmonics (to represent flow) except the $n=1$ component. 
This is illustrated in Fig.~\ref{fig:template}.
Namely, the following function is fit to two-particle correlations in high-multiplicity events,
\begin{equation}
    \frac{dN_\pair^\high}{d\phi} = F\frac{dN_\pair^\low}{d\phi} + \frac{G}{2\pi}\left(1+2\sum_{n=2}^{\infty}V_n\cos n\dphi\right)\,,
    \label{eq:template}
\end{equation}
where $F$ is a fit parameter to scale the low-multiplicity correlations (template).
Note that in the harmonic series, the first harmonic is excluded. In other words, the fitting constraint is to assume that the first harmonic is all nonflow; there is no flow harmonic of the first order.\footnote{Of course, without any imposed constraint,  the fit would not yield any unique result because anything can be described by a Fourier series.}

The template fit assumes, effectively, that the nonflow component in high-multiplicity events scales according to $V_1$. If $V_1$ is inversely proportional to multiplicity (which is approximately the case, see e.g.~Ref.~\cite{Adamczyk:2015xjc}), then the template fit method is similar to the $1/N$ scaling method in Sect.~\ref{sec:scaling}. 
In other words, the assumption here is that nonflow scales with multiplicity in the same way as $V_1$. 
Any deviation of the fit parameter $F$ from the $1/N$ scaling factor can be considered as a modification in nonflow correlations from low to high multiplicity, taken care of by a constant scaling without change in shape. 
This is similar to the near-side jetlike yield-scaled subtraction described in Sect.~\ref{sec:near}, except that the scaling here is determined by the $V_1$ dipole component (mainly the away-side correlation, probably dominated by global momentum conservation~\cite{Borghini:2006yk}) instead of the near-side correlation. 
The template fit can therefore be coined as a $V_1$-scaled subtraction method, just like the near-side scaled subtraction can be recasted as a fitting method to match the near-side yield from small- and large-$\deta$ difference.

Within fit errors, the following equities follow from Eq.~(\ref{eq:template}):
\begin{eqnarray}
    N_\pair^\high &=& F N_\pair^\low + G \,, \label{eq:c0} \\
    N_\pair^\high V_1^\high &=& F N_\pair^\low V_1^\low \,, \label{eq:c1} \\
    N_\pair^\high V_n^\high &=& F N_\pair^\low V_n^\low + G V_n^\sub\,. \label{eq:cn}
\end{eqnarray}
Simple algebra leads to
\begin{equation}
    V_n^\sub =  \left(V_n^\high - \frac{V_1^\high}{V_1^\low} V_n^\low\right) \left/ 
    \left(1 - \frac{V_1^\high}{V_1^\low}\right) \right.\,.
    \label{eq:vntemplate}
\end{equation}
Eq.~(\ref{eq:vntemplate}) is similar to Eqs.~(\ref{eq:scaling}, \ref{eq:near}) except the denominator. The denominator comes about because the normalization $G$ in Eq.~(\ref{eq:cn}) does not include all the pairs in the high-multiplicity events but excludes those genuine nonflow particle pairs. 
It assumes that the nonflow pairs do not participate in the final-state collective flow, and the extracted flow is relative to the reduced average pair multiplicity (the underlying event) excluding those nonflow pairs.
The physics picture is that the high-multiplicity event is composed of nonflow particles (the white area underneath the lower curve in Fig.~\ref{fig:template}) and flow particles (the area between the two curves in Fig.~\ref{fig:template}). 
However, the cross pairs (one from a nonflow source and the other from the collective flow bulk) are included in the normalization $G$ and thus are considered as part of flow. 
This underlying event normalization issue is mostly a small effect but may be relevant for small-system flow studies. To some extent this is an intrinsic uncertainty as it is not a priori clear which picture is more realistic.
This normalization difference is unique to the template fit method. For all other nonflow estimation methods, the final extracted flows are based on the total event (pair) multiplicity, including nonflow particles--see bullet iii) in Sect.~\ref{sec:scaling} (not explicitly stated for other nonflow estimation methods).

The template fit method attempts to account for nonflow changes by scaling the correlations in low-multiplicity events. It is similar in spirit to the near-side jetlike yield-scaled subtraction method,  differing in the assumption of how nonflow changes. So a similar set of assumptions and shortcomings, namely:
\begin{itemize}
    \item[i)] it assumes that the first harmonic is all nonflow and there is no flow $V_1$, so nonflow scales from low to high multiplicity according to the $V_1$ magnitude, without change in the correlation shape;
    \item[ii)] the away-side dijet correlations likely change shape due to jet quenching in medium-to-central heavy-ion collisions, and thus unlikely scale with $V_1$;
    \item[iii)] the near-side nonflow correlations may not scale like $V_1$, and the correlation shape can also change from low- to high-multiplicity collisions.
\end{itemize}

\subsubsection{Dipole scaling method\label{sec:dipole}}
The dipole scaling method~\cite{STAR:2022pfn,STAR:2023wmd} is very similar to the template fit method in Sect.~\ref{sec:template}, also assuming the first harmonic $V_1$ to be all nonflow, and the $V_1$ in low-multiplicity events is scaled to match that in high-multiplicity events, as in Eq.~(\ref{eq:c1}). The $V_n^\sub$ is, however, defined in terms of the total pair multiplicity $N_\pair^\high$, not $G$ as in Eq.~(\ref{eq:cn}). Namely, 
$N_\pair^\high V_n^\high = F N_\pair^\low V_n^\low + N_\pair^\high V_n^\sub$, which leads to 
\begin{equation} \label{eq:dipole}
    V_n^\sub =  V_n^\high - \frac{V_1^\high}{V_1^\low} V_n^\low\,.
\end{equation}
This assumes effectively that all those nonflow particles, while keeping their inter-particle nonflow correlations, have attained collective flow themselves individually. This is as same as the assumption made in Sect.~\ref{sec:scaling} and \ref{sec:near}.
In the template fit method, on the other hand, the low-multiplicity correlated yield is scaled and subtracted first, and those particles are not counted as part of collective flow as aforementioned.
Effectively, the dipole method is as same as the template method except that the subtracted average baseline in the low-multiplicity template is added back to the underlying event.
The dipole method is sometimes called ``$c_1$ method'' (where $c_1\equiv V_1$)~\cite{STAR:2022pfn,STAR:2023wmd}.

\subsection{Data-driven fitting methods\label{sec:fit}}
As seen from Sect.~\ref{sec:deta} and \ref{sec:low}, nonflow contamination is difficult to remove, involving strong assumptions.
Another approach to estimate nonflow contributions is quite different from those described in Sect.~\ref{sec:deta} and \ref{sec:low}. 
It relies on correlation structures observed in data, and identifies local peaks and ridges to attribute to nonflow correlations and the approximately $\deta$-independent underlying distribution to collective flow.
It then involves fitting the correlation data with pre-defined functional forms in a data-driven way.

The advantage of these fitting methods is that they are data driven, having minimal assumptions about the physics of nonflow and no reliance on the evolution of nonflow over centrality/multiplicity. The downside, however, is that the fit functional forms for nonflow correlations are {\em ad hoc} and one needs to ensure that the nonflow functional forms are reasonable descriptions of the data by trial and error, examining fit qualities by using, e.g.~the fit $\chi^2/{\rm ndf}$ values. 

\subsubsection{2D fitting in $(\deta,\dphi)$}\label{sec:2D}
One data-driven approach is to perform two-dimensional (2D) fits to two-particle $(\deta,\dphi)$ correlations~\cite{STAR:2006lbt,STAR:2011ryj,STAR:2022cku,STAR:2023ioo,STAR:2023gzg}. 
Raw two-particle correlations are predominantly of a ``triangle'' shape in $\deta$, which comes from the approximately uniform  single-particle density distribution within limited $\eta$ acceptance (for example, $|\eta|<1$ in the STAR experiment~\cite{Ackermann:2002ad}).
The triangle acceptance can be largely corrected by the mixed-event technique, resulting in two-particle  $(\deta,\dphi)$ correlations, such as the one sketched in Fig.~\ref{fig:2D}, that are dominated by an overall pedestal.
The fine structures atop the pedestal are nonflow correlations and a flow modulation along $\dphi$ (with possibly a $\deta$ dependence). These features have been generally described in the introduction.

The near-side nonflow peaks may be modeled by  2D Gaussians, and the flow modulation is described  by a Fourier series. 
The Fourier coefficients, corresponding to flow harmonics, are typically assumed to be $\deta$ independent. 
The two-particle correlations can then be fit with a functional form including Gaussians and Fourier harmonics.
This 2D fitting method is data-driven; the nonflow correlation shapes are dictated by the data structure and modeled. The method fits the data in a given centrality or multiplicity bin and does not rely on assumptions using low-multiplicity baseline events. 
A recent 2D fit study indicates an approximately 40\% nonflow in central isobar collisions~\cite{STAR:2023ioo,STAR:2023gzg}, in line with the central \AuAu\ data~\cite{Abdelwahab:2014sge} (also see Sect.~\ref{sec:data}) considering the multiplicity dilution of nonflow by approximately a factor of two. 

It is noteworthy that the Fourier flow harmonics can be $\deta$ dependent, for example, due to flow decorrelations~\cite{Bozek:2010vz,Xiao:2012uw,CMS:2015xmx,ATLAS:2017rij}. 
In addition, the $v_n$, although long ranged, can be $\eta$ dependent~\cite{Back:2004mh,STAR:2004jwm,ALICE:2016tlx,CMS:2016est}; such a dependence can result in $\deta$-dependent Fourier coefficients as well.
Including these possible $\eta$-dependencies of the flow harmonics as free parameters in the fit model would not be fruitful as the correlation data would not have sufficient constraining power over these parameters (and those modeling nonflow).
Usually the effects due to the possible $\eta$-dependent flow harmonics are assessed as part of the systematic uncertainties~\cite{STAR:2023ioo,STAR:2023gzg}.

However, the information on the $\deta$-dependencies of flow harmonics can be obtained from other measurements and can be factored into the 2D fit function. We leave discussions on this to the next section.

\subsubsection{1D fitting in $\deta$}
The spirit of the one-dimensional (1D) fitting method is similar to that of the 2D fitting method, except that the $\dphi$ dimension is collapsed into a single number $V_n(\Delta\eta)$ for each harmonic. The $V_n(\Delta\eta)$ is first calculated by Eq.~(\ref{eq:Vn}) in each $\deta$ bin. Then $V_n(\Delta\eta)$ as a function of $\deta$ is decomposed into flow and nonflow through the 1D fitting method. 

\begin{figure}[hbt]
    \includegraphics[width=\linewidth]{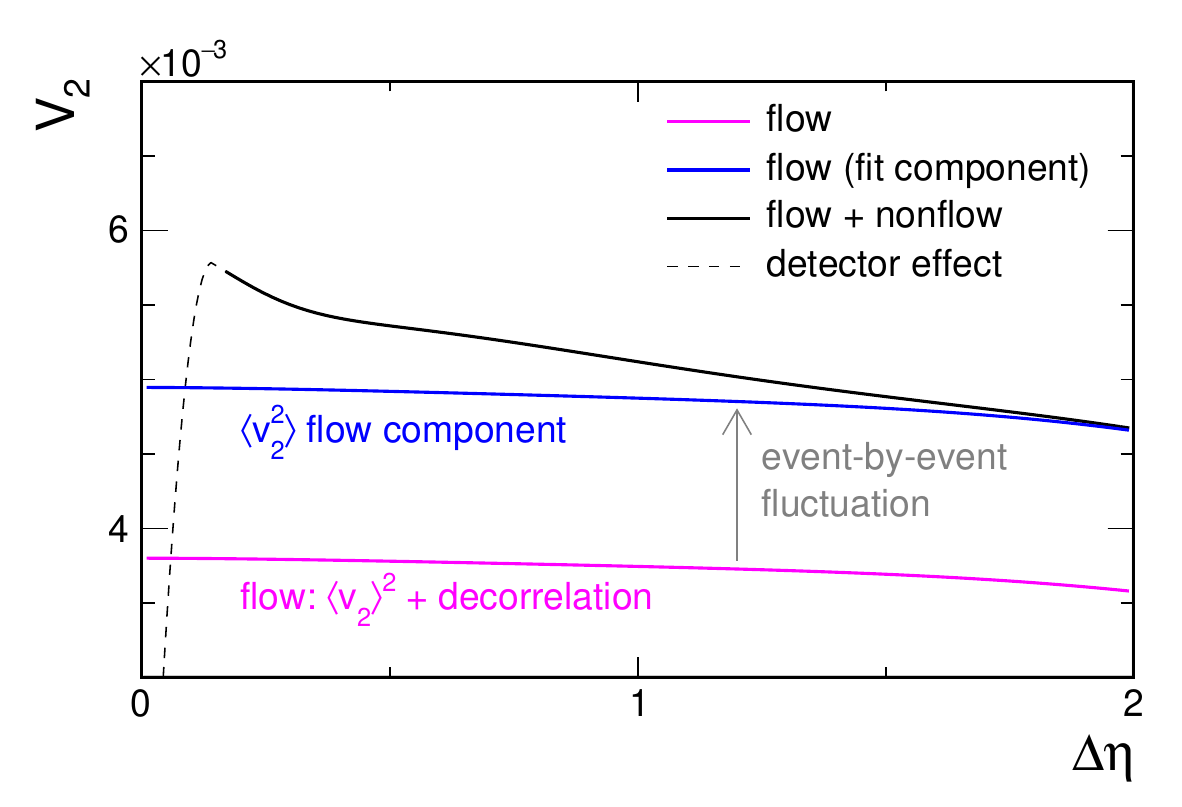}
    \caption{Illustration of the 1D fit method. The black curve indicates a $V_2$ measurement, where the dip at $\deta\sim0$ indicated by the dashed curve is caused by track merging in a typical tracking detector. The magenta curve incorporates $\deta$ convolution from single-particle $v_2(\eta)$ and flow decorrelation $1-F_n\deta$. The blue curve indicates the flow component in $V_2$ scaled up from the magenta curve to include flow fluctuations, with the scaling factor treated as a fit parameter. The difference between the black curve and the blue curve would be nonflow contribution to the $V_2$ measurement. The vertical axis is zero-suppressed and the values are only order-of-magnitude indications.}
    \label{fig:1D}
\end{figure}

Figure~\ref{fig:1D} illustrates a typical $V_2(\deta)$ dependence on $\deta$. 
It is generally a decreasing function of $\deta$, primarily because of the short-range nonflow contributions.
For tracking detectors, track merging/splitting is sometimes important~\cite{STAR:2006lbt,STAR:2011ryj,STAR:2023ioo,STAR:2023gzg}, which creates a dip/peak\footnote{HBT/Bose-Einstein correlations and Coulomb interactions can also create dips and peaks at small pair separation~\cite{Lisa:2005dd}.} at $\deta\sim0$ as indicated by the dashed curve in Fig.~\ref{fig:1D}.
When track merging effects are significant, the average $V_n$ from full-event analysis can  even be smaller than the subevent result which automatically excludes the dip. 
In such a case, the difference between full-event and subevent $V_n$ in the $\deta$-gap methods would not be a good assessment of systematic uncertainties.
On the other hand, when track splitting effects are significant, then the systematic uncertainties would be grossly overestimated.

Compared to the 2D fitting method in Sect.~\ref{sec:2D}, 1D fitting is more straightforward because the $\dphi$ dimension is collapsed in the $V_n$ measurement. 
It is also easier to include the possible $\deta$ dependence of flow and flow fluctuations via $\deta$-dependent $V_n$. 
There are two sources for $\deta$ dependence of flow as aforementioned: 
%
\begin{itemize}
    \item The single particle flow $v_n$ can depend on $\eta$~\cite{Back:2004mh,STAR:2004jwm,ALICE:2016tlx,CMS:2016est}, which would result in a $\deta$-dependent two-particle cumulant $V_n(\deta)$ convoluted from the two single-particle distributions of $v_2(\eta_1)$ and $v_2(\eta_2)$. Such a dependence can be examined, for the case of $n=2$ for example, by  four-particle cumulant measurement $v_{2}\four(\eta)$ where nonflow is largely eliminated, and/or by $v_{2}\zdc(\eta)$ with respect to the first order harmonic plane of spectator neutrons measured by Zero-Degree Calorimeters (ZDC) which is free of nonflow. Note that $v_2\zdc$ measures approximately the average $\mean{v_2}$ and $v_2^2\four\approx\mean{v_2}^2-\sigma^2$ where $\sigma$ denotes Gaussian width of $v_2$ fluctuations, whereas the flow component in the two-particle measurement $V_2$ is $v_2^2\two=\mean{v_2}^2+\sigma^2$~\cite{Poskanzer:1998yz}. As an approximation, one may obtain $v_2\two(\eta)$ from $v_{2}\four(\eta)$ and $v_{2}\zdc(\eta)$. Alternatively, one may  make the reasonable assumption that the effect of $v_2$ fluctuations is proportional to $\mean{v_2}$ and apply a scale factor as a free fit parameter to $v_{2}\four(\eta)$ or $v_{2}\zdc(\eta)$ to obtain $v_{2}\two(\eta)$. 
    \item Another source of $\deta$ dependence is flow decorrelation, $r_{n}(\deta) \approx 1 - F_{n} \deta$~\cite{Bozek:2010vz,Xiao:2012uw,CMS:2015xmx,ATLAS:2017rij}. 
    Such decorrelations can result from fluctuations of the harmonic flow magnitude as well as harmonic planes over $\deta$~\cite{Bozek:2018nne,Bozek:2021mov}. The decorrelation parameter $F_n$ of a few percent has been measured at RHIC~\cite{Yan:2023ugh} and the LHC~\cite{CMS:2015xmx,ATLAS:2017rij}. 
\end{itemize}
The $\deta$ dependence of the flow component in two-particle $V_2(\deta)$ is the combination of the two effects above. This is illustrated by the magenta curve in  Fig.~\ref{fig:1D}. 

The nonflow component in $V_2(\deta)$ can be modeled by typical functional forms like Gaussians in $\deta$.
Fits are applied to $V_2(\deta)$ excluding the track merging $\deta\sim0$ region, or alternatively including a negative Gaussian at $\deta=0$ to model it.
The fit function, schematically, is
\begin{equation}\label{eq:fit}
    V_2(\deta) = \mean{v_2}^2\mbox{-shape} \times (1-F_n\deta) \times f + \mbox{nonflow}.
\end{equation}
The blue curve in Fig.~\ref{fig:1D} is an illustration of the fitted flow, scaled up  from the shape given in magenta. 
This scaling factor, the $f$ in Eq.~(\ref{eq:fit}), is the effect of flow fluctuations, assumed to be independent of $\deta$ and treated as  a fit parameter.
The idea of the fit is to have the best description of $V_2(\deta)$ using the known shape of flow and flow fluctuations, and deciding by trial and error on the functional forms for the $\deta$ dependence of nonflow. 

The away-side dijet correlations are always the most notorious nonflow to handle. If the away-side correlations in $\deta$ are just like flow correlations such that both are relatively uniform, then 1D fitting cannot distinguish them. One may resort to 2D fitting if the away-side correlations are relatively sharp-peaked in $\dphi$ such that one is reasonably confident that it is nonflow. If it is broadly distributed, then there is really no way to distinguish it from flow as Fourier components can describe any functional shape. However, it is more likely that the away-side correlations in $\deta$ differ from flow; for instance, away-side jet correlations could be more peaked at $\deta=0$ at relatively high $\pt$ (at the highest-$\pt$ extreme the dijet will have to be both at midrapidity), or at low $\pt$ perhaps dipped at $\deta=0$ (e.g.~because of longitudinal flow). 
%
On the other hand, if the away-side correlations in $\deta$ are sufficiently different from flow, then it is possible to distinguish it in 1D fitting by examining the data structure in $V_2(\deta)$.


It is noteworthy that the 1D fitting result and the $\deta$-gap results are certainly related because they both come from the data. However, the flow components extracted from these two methods differ because of the different assumptions involved. The flow component from the $\deta$-gap method is whatever is measured in the $V_2$ beyond the $\deta$ gap, while that from the 1D fitting method is more sophisticated, dependent of the flow shape and decorrelation in $\deta$. Nevertheless, one can calculate/integrate from the 1D fitting result the flow component that would be extracted with the $\deta$-gap method. 

\subsubsection{Symmetry method in $(\eta_1,\eta_2)$}
STAR performed another data-driven analysis by examining $V_2(\eta_1,\eta_2)$ as a function of the two particles' pseudorapidities~\cite{Abdelwahab:2014sge}. 
This method~\cite{Xu:2012ue} exploits the $\eta$ reflection symmetry in symmetric heavy-ion collisions by comparing two pairs of pseudorapidity bins, one at $(\eta_1,\eta_2)$ and the other at $(\eta_1,-\eta_2)$, and thus does not assume any particular shapes for flow as functions of $\eta$. 
Any difference between the two pairs must arise from $\deta$-dependent physics.
The $\deta$-dependent component and the $\deta$-independent component can be  identified in $V_2(\eta_1,\eta_2)$. 
The former is associated with the combination of nonflow and $\deta$-dependent flow fluctuations (decorrelation). 
The $\deta$-independent part is associated with flow plus $\deta$-independent flow fluctuations. They are  found to be independent of $\eta$ within the limited STAR acceptance of $|\eta|<1$. 
STAR has also supplemented the analysis with the four-particle cumulant $V_2\four(\eta_1,\eta_2)$ as a function of the particles' pseudorapidities, 
albeit large uncertainties, to separate flow fluctuations from the average flow magnitude.

This method has the least model assumption and the functional dependencies on $\eta$ and $\deta$ are determined from data. 
The disadvantages of the method are the required high statistics,\footnote{The four-particle cumulant $V_2\four(\eta_1,\eta_2)$ is also analyzed in two dimensions and is limited by statistics with  large uncertainties.} and the reduction of data to  functional forms to economically describe the various components of flow and nonflow requires iterations of trial and error.
The estimated nonflow in the most central 0--5\% \AuAu\ collisions from this method is on the order of 20\% with a relatively large uncertainty~\cite{Abdelwahab:2014sge}.

\section{An example case study}\label{sec:data}

As an example, we illustrate the various nonflow subtraction methods using  STAR data of dihadron $(\deta,\dphi)$  correlation functions in \AuAu\  collisions at $\snn=200$~GeV published in Ref.~\cite{STAR:2011ryj}. The  data are presented in the form of $\Delta\rho/\sqrt{\rho_{\rm ref}}=\sqrt{\rho_{\rm ref}}(r-1)$ where $r$ is the $(\deta,\dphi)$ correlation function divided by the properly normalized mixed-event one, and $\sqrt{\rho_{\rm ref}}=dN/d\eta/2\pi$.

{\em The $\deta$-gap method.}
One can calculate $V_n$ from the correlation function $r(\deta,\dphi)$ by Eq.~(\ref{eq:Vn}) with certain minimum $\deta$ gaps.
The values of $V_2$ for the top 0--5\% centrality are listed in Table~\ref{tab:data} for various $\deta$ gaps, together with the relative reduction from the inclusive one (i.e.~no $\deta$ gap requirement). This reduction corresponds to the removed nonflow fraction by each $\deta$ gap. 

It is noteworthy that the correlation functions have already been corrected for the triangle $\deta$ acceptance via the mixed-event technique~\cite{STAR:2011ryj}. 
The calculated numerical value of $V_n$ is dependent of this acceptance correlation because nonflow is primarily composed of short-range correlations, decreasing with increasing $\deta$. 
Because flow is approximately $\eta$ independent, the $\deta$-acceptance is sometimes uncorrected in experimental analysis of two-particle cumulant $V_2$, e.g., in early works of Ref.~\cite{STAR:2002hbo,Adams:2004bi}. 
Without the triangle acceptance correction, the calculated $V_2$ values would be larger. 
For numerical illustration, we ``uncorrect'' the STAR data by the perfect triangle acceptance, and tabulate the obtained $V_2$ values in Table~\ref{tab:data} together with the corresponding nonflow reductions. In this case the nonflow effects are larger.

\begin{table*}
\caption{An example case study of nonflow estimations by various methods in top 0--5\% \AuAu\ collisions at $\snn=200$~GeV using midrapidity $(\deta,\dphi)$ correlation data from STAR~\cite{STAR:2011ryj}. Quoted values are two-particle cumulant $V_2$ ($\times 10^{-4}$) with statistical uncertainties. The midrapidity multiplicity densities are $dN/d\eta=5.2$, 13.9, and 671 for 84--93\%, 74--84\% and 0--5\% centralities, respectively. The data are corrected for $\deta$ acceptance via mixed events. The last column ``uncorrects'' the data using the perfect triangle acceptance.}
\label{tab:data}
\centering
\begin{tabular}{C{2cm}R{3cm}C{0.2cm}L{3cm}L{1.5cm}R{1cm}C{1cm}L{1.5cm}R{1cm}}
\hline 
Method & \multicolumn{3}{c}{Implementation} & \multicolumn{2}{l}{$\deta$-acc.~corrected} && \multicolumn{2}{l}{$\deta$-acc.~uncorrected} 
\\ \hline 
\multirow{3}{*}{$\deta$-gap}
        & \multicolumn{3}{c}{---}              & $6.12\pm0.05$  &           && $6.64\pm0.03$    &         \\ 
        & \multicolumn{3}{c}{$|\deta|>1.04$}   & $5.30\pm0.10$  & $-13\%$   && $5.45\pm0.07$    & $-18\%$ \\
        & \multicolumn{3}{c}{$|\deta|>1.52$}   & $5.08\pm0.18$  & $-17\%$   && $5.25\pm0.14$    & $-21\%$ \\ 
\hline 
\multirow{10}{*}{\parbox{2cm}{Low-multiplicity subtraction}}
    &\multirow{2}{*}{$1/N$ scaling}&& 84--93\% & $5.67\pm0.06$  & $-7\%$    && $6.07\pm0.04$    & $-9\%$ \\
                        &          && 74--84\% & $5.13\pm0.06$  & $-16\%$   && $5.42\pm0.03$    & $-18\%$ \\
    & \multirow{2}*{$1/N$ scaling} && 84--93\% & \multirow{2}*{$5.10\pm0.08$}& \multirow{2}*{$-4\%$} && \multirow{2}*{$5.25\pm0.08$} & \multirow{2}*{$-4\%$} \\
                        &          && $|\deta|>1.04$ &          &           && & \\     
    & \multirow{2}*{dipole scaled} && 84--93\% & \multirow{2}*{$5.04\pm0.11$}& \multirow{2}*{$-5\%$} && \multirow{2}*{$5.21\pm0.09$} & \multirow{2}*{$-4\%$} \\
                        &          && $|\deta|>1.04$ &          &           && & \\ \\
                & near-side scaled && 84--93\% & $4.93\pm0.13$  & $-19\%$   && $5.12\pm0.13$    & $-23\%$ \\ 
\hline 
\multirow{3}{*}{1D fit}
    & \multicolumn{3}{c}{exp. + const}         & $5.23\pm0.15$  & $-15\%$   && & $-21\%$ \\
    & \multicolumn{3}{c}{Gaus. + const}        & $5.48\pm0.08$  & $-10\%$   && & $-17\%$ \\
    & \multicolumn{3}{c}{two-Gaus. + const}    & $5.29\pm0.12$  & $-14\%$   && & $-20\%$ \\
\hline
\end{tabular}
\end{table*}


\begin{figure*}[hbt]
    \begin{minipage}{0.5\linewidth}
    \xincludegraphics[width=0.9\linewidth,label=\hspace*{0.5cm}a),fontsize=\Large]{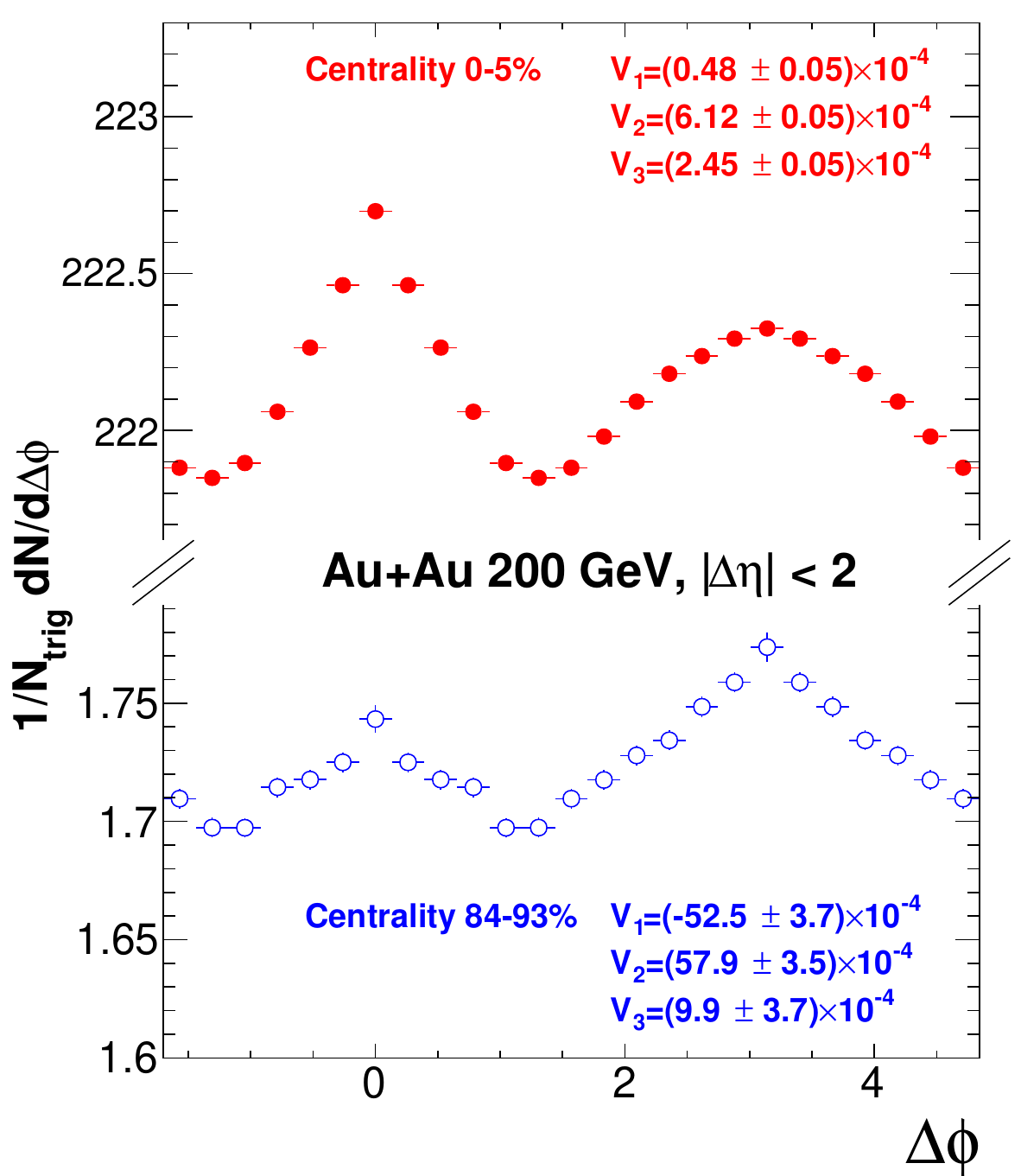}
    \end{minipage}
    \begin{minipage}{0.4\linewidth}
    \xincludegraphics[width=0.8\linewidth,label=\hspace*{0.1cm}b),fontsize=\large]{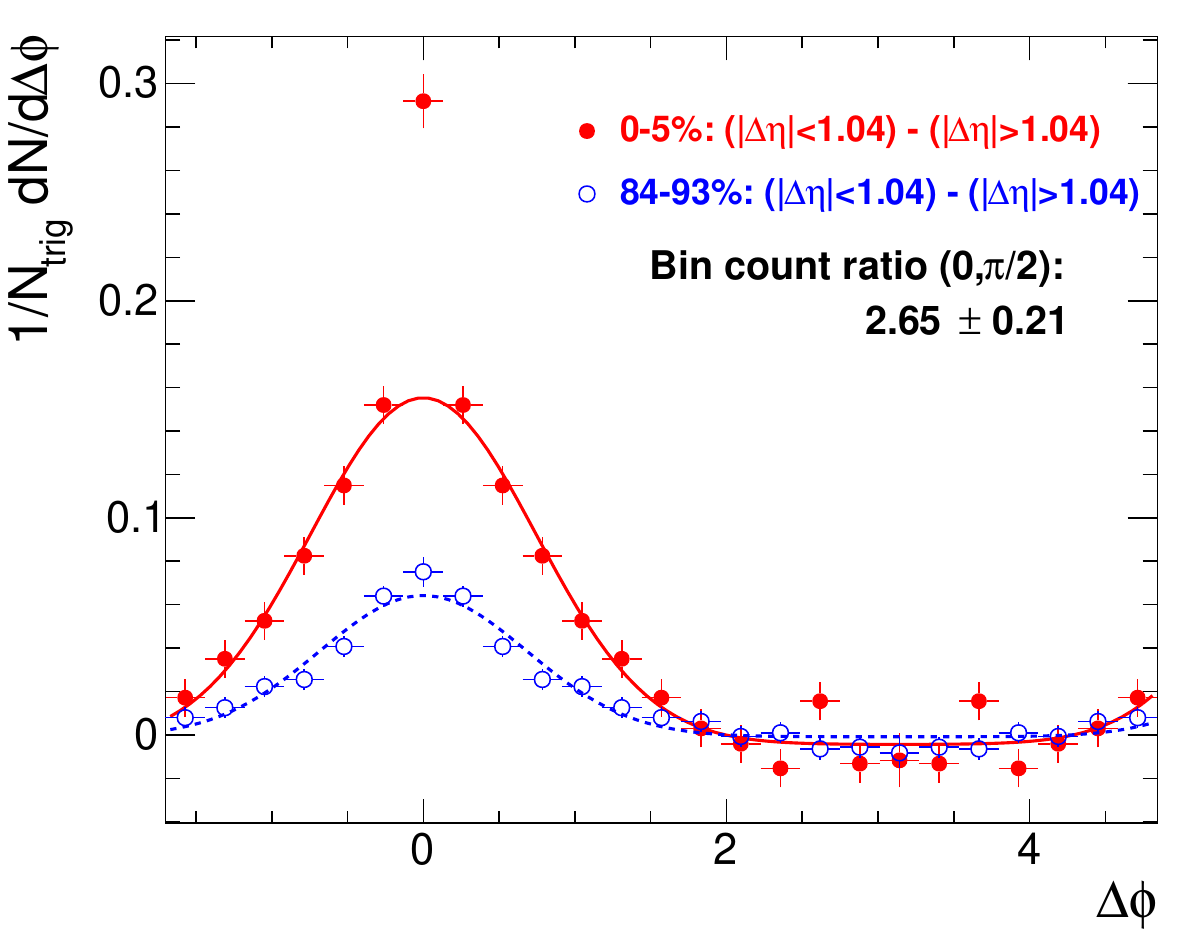}
    \xincludegraphics[width=0.8\linewidth,label=\hspace*{0.1cm}c),pos=nwlow,fontsize=\large]{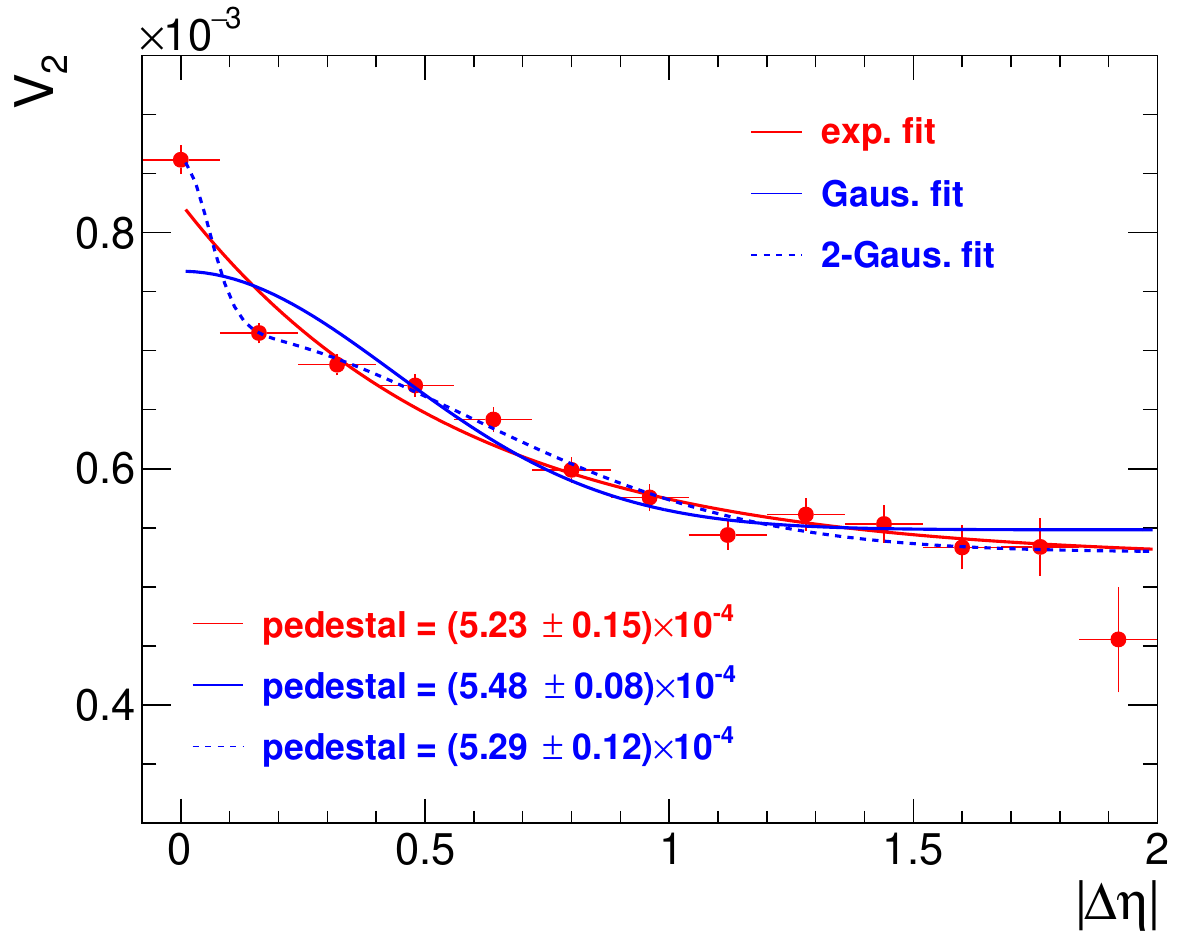}
    \end{minipage}
    \caption{(a) Per-hadron normalized $\dphi$ correlations  in central 0--5\% and peripheral 84--93\% \AuAu\ collisions at $\snn=200$~GeV. The Fourier coefficients are written on the plots.
    (b) $\dphi$ correlation function at $|\deta|<1.04$ minus that at $1.04<|\deta|<2$ after properly normalized on the away side ($\pi/2<\dphi<3\pi/2$). The Gaussian curves are to guide the eye. 
    (c) $V_2(\deta)$ as a function of $\deta$ of particle pair. Fit functions are superimposed to model nonflow correlations atop of a pedestal,  the underlying flow. 
    The $\dphi$ correlation data points are symmetrized about $\dphi=0$ and $\pi$. Error bars are statistical uncertainties.
    Data from Ref.~\cite{STAR:2011ryj} by the STAR Collaboration.}
    \label{fig:data}
\end{figure*}

{\em The $1/N$ scaling method.} The per-hadron normalized correlated yields, $\frac{1}{N_{\rm trig}}\frac{dN}{d\dphi}$, can be obtained from projections of $\sqrt{\rho_{\rm ref}}\times r$~\cite{STAR:2011ryj}. These correlated yield distributions are shown in Fig.~\ref{fig:data}(a) for the most central 0--5\% and most peripheral 84--93\% collisions. The first three Fourier coefficients are written on the corresponding plot. Using the simple $1/N$ scaling method of Eq.~(\ref{eq:scaling}), one can readily obtain the ``nonflow-subtracted'' flow anisotropy, $V_n^\sub$. The values of elliptic flow $V_2^\sub$ for the top 0--5\% centrality are tabulated in Table~\ref{tab:data}. As discussed in Sect.~\ref{sec:simple}, the result depends on what centrality is used as the low-multiplicity event class. The $V_2^\sub$ calculated using the 74--84\% centrality (instead of the 84--93\% one) for low-multiplicity subtraction is also tabulated in Table~\ref{tab:data}. The corresponding nonflow reduction is larger, as one would expect. 
The effects of the $\deta$ triangle acceptance are also shown in Table~\ref{tab:data}.

In this low-multiplicity subtraction exercise we have used  pair correlations without $\deta$ gap. 
One can certainty use the correlations with a minimum $\deta$ requirement and then further apply low-multiplicity subtraction, as we show in Table~\ref{tab:data} as well. The relative nonflow reduction is smaller because the base $V_2$ with the $\deta$ gap has a significant portion of nonflow already removed. However, the nonflow reduction is finite, indicating  remaining nonflow beyond the applied $\deta$ gap. 
Note that the $\deta$-acceptance effect is in this case negligible because the nonflow contribution at large $\deta$ is relatively small and approximately $\deta$ independent.

We may repeat the exercise using the directly measured $v_2$ values from STAR~\cite{Adams:2004bi} (the triangle acceptance is not corrected in these data).
The measured $v_2$ in peripheral 70--80\% and top 0--5\% central \AuAu\ collisions are $v_2\two^{\rm peri}\approx6.9\%$ and $v_2\two^{\rm cent}\approx2.4\%$, respectively~\cite{Adams:2004bi}. Assuming the former is all nonflow, the $1/N$ scaling of nonflow would result in a nonflow of $(v_2^{\rm peri})^2\frac{\Nch^{\rm peri}}{\Nch^{\rm cent}} \approx 1.7\times10^{-4}$, i.e.~30\% nonflow in $(v_2^{\rm cent})^2$ of the latter. 
Choosing 90-100\% centrality as the low-multiplicity baseline would yield a significantly smaller nonflow estimate~\cite{Jia}. 
As discussed in Sect.~\ref{sec:low}, it is possible that the 90-100\% (or 84-93\%) centrality, which is defined by the event multiplicity, is biased towards too soft interactions, thus  underestimating the nonflow to be used as subtraction for central collisions. On the other hand, the higher 70-80\% centrality data  likely contain flow, and  thus the 30\% estimate is likely an overestimate of nonflow.
It is subject to the uncertainty what peripheral range best represents the underlying nonflow magnitude of those central collision events. 

To avoid the peripheral multiplicity bias, minimum bias (MB) \pp\ collisions may be a better reference for nonflow subtraction.
STAR has measured accumulative correlation as a function of $\pt$, $\mean{\sum_i\cos2(\phi_{\pt}-\phi_i)}$, where $\phi_{\pt}$ is the azimuthal angle of the particle of interest from a given $\pt$ bin and $\phi_i$ is that of the reference particle. 
The flow component in the accumulative correlations is $Nv_2(\pt)\bar{v}_2$, and the nonflow contribution would be constant under the $1/N$ scaling assumption so the ratio \pp/\AuAu\ of the accumulative correlations can be directly related to the nonflow fraction in \AuAu\ collisions. 
The accumulative correlations are presented as functions of $\pt$ in \pp\ and \AuAu\ collisions (Fig.~1 in Ref.~\cite{STAR:2004amg}). The peripheral 80--100\% \AuAu\ collisions are comparable to \pp, suggesting that these peripheral collisions are dominated by nonflow.
In the top 0--5\% central collisions, the measured ratio is on average 15\%, or 12\% after correcting for the occupancy-dependent detector efficiency of 20\% from \pp\ to central \AuAu~\cite{Abelev:2008ab}.
One may consider that the mean $\mean{\pt}$ in central \AuAu\ collisions is larger than in \pp, and because the accumulative correlation is an increasing function of $\pt$, the $N$-weighted \pp/\AuAu\ ratio (or the ratio of the accumulative correlations at the corresponding $\mean{\pt}$ values) is smaller, approximately 8\%~\cite{Jia}.
On the other hand, one may argue that the \pp/\AuAu\ ratio should be taken at higher $\pt$ for \pp\ than for \AuAu\ because of jet energy loss in the latter, which would yield a larger ratio.
These considerations suggest that nonflow estimations by the $1/N$ scaling, even without multiplicity biases, still have large uncertainties because of strong assumptions made in this method.

{\em The dipole scaling method.}
With the Fourier coefficients from peripheral and central collisions, we can readily apply the dipole scaling method of Eq.~(\ref{eq:dipole}). It is generally expected that the negative dipole is approximately inversely proportional to $N$, so the $1/N$ scaling and the dipole scaling would give similar results of $V_n^\sub$.
Obviously, the correlations shown in Fig.~\ref{fig:data}(a) without $\deta$ gap is not suitable for the dipole scaling method because the dipole $V_1$ in central collisions is not even negative because of the large near-side nonflow peak at $\dphi\sim 0$.
Instead, we apply the dipole scaling method to the correlations with a minimum $\deta$ gap of 1.04 (the bin edge of a $\deta$ bin). The $V_1$ values for $\dphi$ correlations in $1.04<|\deta|<2$ are $(-1.40\pm0.07)\times10^{-2}$ and $(-1.30\pm0.10)\times10^{-4}$ for 84--93\% and 0--5\% centralities, respectively. The $V_1$ ratio of $0.0093\pm0.0009$ compares reasonably well to the inverse multiplicity ratio of $0.0077$.
The $V_2^\sub$ by Eq.~(\ref{eq:dipole}) is shown in Table~\ref{tab:data}.
Again the nonflow reduction is small because the base $V_2$ with the $\deta$ gap  is already void of significant nonflow.

The template fit method is not shown in Table~\ref{tab:data}. It is a close variation from the dipole scaling method as discussed in Sect.~\ref{sec:template} and Sect.~\ref{sec:dipole}.
The difference between Eqs.~\ref{eq:vntemplate} and \ref{eq:dipole} is the $V_1$ ratio in the denominator, which is only on the order of 1\%.

{\em The near-side scaling method.}
Figure~\ref{fig:data}(b) shows the per-hadron normalized  correlated yield difference in $\dphi$ between small- and large-$\deta$ ranges, $|\deta|<1.04$ and $1.04<|\deta|<2$, respectively. A slight normalization adjustment is applied before  taking the difference, a couple of percent for peripheral collisions and negligible for central collisions, such that the away-side ($\pi/2<\dphi<3\pi/2$) correlated yield is averaged to zero. 
This correlation difference represents the near-side short-range nonflow correlations under the assumption of $\deta$-independent collective flows.
The ratio of the near-side correlated yield in central collisions over that in peripheral collisions is $Y_\high/Y_\low=2.65\pm 0.21$,\footnote{The $\dphi=0$ data point in central collisions may be affected by track splitting so the ratio is taken excluding the $\dphi=0$ data points.} and this ratio is applied in the near-side subtraction method of Eq.~(\ref{eq:near}) to the scaling of nonflow in addition to the simple multiplicity dilution.
The near-side scaled nonflow subtracted $V_2^\sub$ value is shown in Table~\ref{tab:data}. 
The nonflow effect is significantly larger than the simple $1/N$ scaling because the near-side correlated yield is significantly larger in central collisions. 


{\em The 1D fitting method.}
The $V_2$ as a function of $|\deta|$ is shown in Fig.~\ref{fig:data}(c). The $V_2$ value with a certain minimum $\deta$ gap requirement is simply the pair density-weighted integral of this plot over the range beyond the given $|\deta|$ gap size. Take the $\deta$-gap idea a step forward would be the 1D fitting method. For good accuracy, one would want to take into consideration other flow measurements that are less vulnerable to nonflow, such as those from  ZDC measurements or multi-particle cumulants as discussed in Sect.~\ref{sec:fit}.
For the   illustration purpose here, we assume simply that the flow is independent of $\deta$ and use various functional forms for nonflow correlations. Specifically, we use exponential, single Gaussian, and double Gaussian functions. The fits are superimposed in Fig.~\ref{fig:data}(c). The fit pedestals, corresponding to the ``real'' flow, are written on the plot, and tabulated in Table~\ref{tab:data} with the corresponding nonflow reduction.
The flow for the $\deta$-acceptance uncorrected data would be the same as the fit pedestal. 
The corresponding nonflow reduction is larger as shown in Table~\ref{tab:data}, because the inclusive $V_2$ measurement is larger containing stronger nonflow contribution. 

It is clear from our example case study that the various nonflow subtraction methods result in a wide range of nonflow estimations. 
The nonflow contamination in the top 0--5\% central \AuAu\ collisions from this illustrative exercise is approximately 20\%.
The absolute range in the nonflow estimation and thus the relative uncertainty on the flow result is on the order of 10\%. 
The nonflow effect is expected to be smaller at the LHC than at RHIC because of the larger multiplicity, thus larger dilution of nonflow.
A significant fraction of nonflow may be removed by a large $\deta$ gap. 
This is in line with findings at the LHC~\cite{ALICE:2010suc,ALICE:2016ccg} and RHIC~\cite{STAR:2010ico}. 
However, nonflow is not completely absent at $|\deta|>1$. Low-multiplicity subtractions and fitting methods indicate an additional a few percent (absolute) nonflow contamination beyond $|\deta|>1$.

\section{Discussions and Summary}
Nonflow includes all few-body correlations in a collision event except the global flow correlations where all particles are correlated over the entire event. Those nonflow correlations cannot possibly be fully measured, and various estimation/subtraction methods have been devised, as reviewed here in a single place. They can be  broadly divided into three categories: (1) $\deta$-gap, (2) low-multiplicity subtraction, and (3) data-driven fits. All of them have assumptions, some of which are strong and some are less so; they come with different pros and cons. 
Because of the various degrees of assumptions, assessments of systematic uncertainties on nonflow are challenging. 
We summarize in Table~\ref{tab} the various nonflow estimation methods regarding their assumptions, pros, cons, and the  sources of uncertainties involved.
\begin{table*}
    \caption{Summary of the assumptions, pros, cons, and sources of uncertainties involved in the various nonflow estimation methods. 
    }
    \label{tab}
    \centering
    \begin{tabular}{L{1.7cm}L{2.4cm}L{3.4cm}L{1.8cm}L{2.7cm}L{5cm}}
    \hline 
    Method & Implementation & Assumptions & Pros & Cons & Uncertainty sources \\
    \hline
    \myvcell{$\deta$-gap} & 
    \myvcell{
    \begin{enumerate}[leftmargin=*]
        \item pair $\deta$ 
        \item two-subevent 
        \item multi-subevent
    \end{enumerate}
    } &
    \myvcell{
    \begin{itemize}[leftmargin=*]
        \item[-] nonflow is short-ranged
    \end{itemize}
    } & 
    \myvcell{simple} & 
    \myvcell{difficult to assess systematic uncertainties} & 
    \myvcell{
    \begin{itemize}[leftmargin=*]
        \item[-] choice of $\deta$ gap size (nonflow beyond $\deta$)
        \item[-] flow dependence on $\eta$ 
        \item[-] flow decorrelation in $\deta$
    \end{itemize}
    } \\ \hline
    \myvcell{Low-multiplicity subtraction} & 
    \myvcell{
    \begin{enumerate}[leftmargin=*]
        \item $1/N$ scaling
        \item near-side scaling
        \item template fit 
        \item dipole scaling
    \end{enumerate}
    } & 
    \myvcell{
    \begin{itemize}[leftmargin=*]
        \item[-] low-multiplicity events have no flow
        \item[-] certain multiplicity dependencies for nonflow
        \item[-] template fit differs from dipole scaling only in base normalization
        \end{itemize}
    } & 
    \myvcell{restrained by low-multiplicity correlations} & 
    \myvcell{strong model dependencies; near-side correlation analysis can be difficult at low $\pt$} & 
    \myvcell{
    \begin{itemize}[leftmargin=*]
        \item[-] choice of low-multiplicity events
        \item[-] degree of validity of assumptions (nonflow shape can change, near- and away-side may scale differently, nonflow sources may not scale with multiplicity)
        \item[-] multiplicity selection biases, especially in small systems
        \item[-] detector effects (track merging/splitting) is multiplicity dependent
    \end{itemize}
    } \\ \hline
    \myvcell{Data-driven fitting} & 
    \myvcell{
    \begin{enumerate}[leftmargin=*]
        \item 2D fit
        \item 1D fit
        \item $\eta$-symmetry
    \end{enumerate}
    } & 
    \myvcell{
    \begin{itemize}[leftmargin=*]
        \item[-] flow and nonflow shapes are different
    \end{itemize}
    } & 
    \myvcell{least model dependent} & 
    \myvcell{fitting can be demanding (trial and error)} & 
    \myvcell{
    \begin{itemize}[leftmargin=*]
        \item[-] functional forms for nonflow correlations
        \item[-] $\deta$ dependence of flow and flow fluctuations (which can be input from elsewhere)
    \end{itemize}
    } \\ \hline
    \end{tabular}
\end{table*}
\begin{enumerate}
    \item The $\deta$-gap methods, while simple to implement, are not clean. The interpretation of  results of a given $\deta$-gap analysis is subject to issues like the $\eta$-dependence and $\deta$-decorrelation of flow. The systematic uncertainties are hard to quantify; comparing to full-event (no $\deta$ gap) benchmark can, for example, be bitten by the usual track-merging/splitting detector artifacts. An improvement would be to examine the results as functions of the applied $\deta$-gap size, which would constitute the data-driven 1D fitting method.
    \item The low-multiplicity subtraction methods come with the large uncertainty in the assumptions of the multiplicity/centrality dependence of nonflow    and the arbitrariness in the choice of low-multiplicity event class. 
    Strictly confining within a given set of assumptions, one may arrive at a nonflow estimate with relatively small systematic uncertainty. However, this is only valid when the assumptions are correct; 
    more likely, the estimated nonflow is systematically biased, and loosening the assumptions would yield a wide range of uncertainties. 
    \item The data-driven fitting methods are probably the best at our disposal with the fewest assumptions and least model dependency. However, it requires attentive examination of the data and identification of nonflow correlation shapes. The fitting can be demanding and requires thorough inspection.
\end{enumerate}

Naturally, different methods are different in their degrees of efficacy in removing nonflow. 
The $\deta$-gap method is used in essentially all modern flow analyses. The low-multiplicity subtraction methods have been actively explored in data analyses, especially in the context of small-system collisions. 
The fitting methods take more efforts; the methods are not new but are less explored.
We have included an example case study of the various nonflow subtraction methods using the top 0--5\% central \AuAu\ data at $\snn=200$~GeV published by STAR~\cite{STAR:2011ryj}. 

In general, nonflow contamination is severe in peripheral collisions and become less so towards more central collisions. However, since the elliptic flow also decreases with increasing centrality because of the more spherical collision zone, nonflow contamination in central collisions can still be appreciable, whereas it is generally the smallest in midcentral collisions.
In the top 0--5\% central \AuAu\ collisions at RHIC, for example, the data-driven fitting methods indicate a nonflow fraction of 20\%, with a typical relative systematic uncertainty of 20\%~\cite{Abdelwahab:2014sge,Feng:2021pgf,STAR:2023gzg,STAR:2023ioo}.
From low-multiplicity subtraction, one can get a wide range of nonflow estimate, probably 10--30\%, depending on assumptions of multiplicity dependence of nonflow and what low-multiplicity events are taken as nonflow baseline.
Multiplicity dependence of nonflow involve two factors, one is a possibly faster increase in the number of nonflow sources than the multiplicity, and the other is a possible change in the correlation of each nonflow source with multiplicity/centrality.

Nonflow subtraction is particularly challenging in small system collisions to search for signs of collectivity in high-multiplicity events. The reasons are many-fold: the nonflow contamination is high and likely dominates the $V_2$ measurement, multiplicity selection biases are significant for both low- and high-multiplicity events, how nonflow varies from low- to high-multiplicity collisions is largely unknown (partially because of the selection biases). The task is relatively easier at the LHC than at RHIC because of the larger detector longitudinal acceptances,  the larger multiplicities produced,  and  the more likelihood to have collectivity at the higher energies of LHC. At RHIC the analysis is significantly more difficult; for example, the difference between PHENIX~\cite{PHENIX:2018lia,PHENIX:2021ubk} and STAR~\cite{STAR:2022pfn,STAR:2023wmd} on flow in $p$+Au, $d$+Au, and $^3$He+Au is not fully settled.

One may naively expect that some of the methods would at least give the lower limit of nonflow (thus upper limit of flow). For example, the $\deta$-gap method could give a lower limit because one expects residual nonflow to still contribute beyond a certain $\deta$ gap. This is true if one confines to the physics interpretation of flow and nonflow within the measured $\deta$ region. If one infers the $\deta$-gap measurement of flow to, e.g.~the entire $|\deta|>0$ region, then the measured flow at large $\deta$ can be lower than the true flow over the entire $\deta$ range because flow and/or flow fluctuation effects can decrease with $\deta$. 
This could then be a lower limit of flow and upper limit of nonflow; however, this is really a physics interpretation issue.

The difficulties in estimating/subtracting nonflow reflects the fact that nonflow cannot be fully and thoroughly measured and the physics evolution of nonflow with collision system and centrality/multiplicity is not well understood.
It is therefore important to examine various estimation/subtraction methods, when strong assumptions are involved, in order to arrive at a robust nonflow estimate with faithful systematic uncertainties, which often takes the majority effort of data analysis. This is particularly important when data are compared to theoretical models, such as hydrodynamics, where nonflow effects are not fully incorporated, to draw quantitative physics conclusions. 

We end our topical review of nonflow by a set of recommendations:
\begin{itemize}
    \item Use multiple methods (as many as feasible), possibly combine methods, and assess the range of uncertainties in the results;
    \item Always strive to use data-driven methods even though they may be more demanding;
    \item Employ information and/or constraints from other measurements as much as possible;
    \item Use multi-particle cumulants wherever feasible, keeping in mind the different flow fluctuation effects in two- and multi-particle cumulants;
    \item For small systems, take extreme care in assessing the possibly severe multiplicity biases in event selections.
\end{itemize}
We hope our review summarizing the various nonflow estimation methods in a single place is helpful to future researches.

\section*{Acknowledgment} 
We thank Dr.~Zhenyu Chen for organizing the {\em 4th International Workshop on QCD Collectivity at the Smallest Scales}, which stimulated the writing of this work.
F.W.~expresses his gratitude to Dr.~Sergei Voloshin and Dr.~Jurgen Schukraft for many fruitful discussions.
This work is supported in part by the U.S.~Department of Energy (Grant No.~DE-SC0012910).

\bibliography{./ref}

\end{document}